\documentclass[12pt]{iopart}

\usepackage{amsfonts}
\usepackage{latexsym}

\begin{document}
\input{epsf.tex}
\begin{center}\Large The XX--model with boundaries.\\ 
Part III: Magnetization profiles and boundary bound states
\\[1cm]
\normalsize
Ulrich Bilstein\footnote{E-mail address: \tt bilstein@th.physik.uni-bonn.de}\\[0.5cm]
 Universit\"{a}t Bonn \\
                    Physikalisches Institut, Nu\ss allee 12,
                    D-53115 Bonn, Germany\\[1.3cm]
\end{center} 
{\bf \small Abstract.}
\small
We calculate the magnetization profiles of the $\sigma_j^x$ and $\sigma_j^z$ operators
  for the XX-model with hermitian boundary terms. We study the profiles on the finite 
chain and in the continuum limit. The results are discussed in the context of conformal 
invariance. 
We also discuss boundary excitations 
and their effect on the magnetization profiles.
\\[0.2cm]
\normalsize
\section{Introduction}

We continue our study of the XX-model with boundaries defined
by the Hamiltonian
\begin{equation}
\fl
H=\frac{1}{2}\sum_{j=1}^{L-1}
\left[\sigma_j^+\sigma_{j+1}^-+\sigma_j^-\sigma_{j+1}^+\right]
+\frac{1}{\sqrt{8}}\left[\alpha_-\sigma_1^-+\alpha_+\sigma_1^+
+\alpha_z\sigma_1^z+\beta_+\sigma_L^++\beta_-\sigma_L^-+\beta_z\sigma_L^z\right]
\label{HXX}
\end{equation}
where $\sigma^{\pm}_j=\frac{1}{2}( \sigma_j^x \pm \rmi \sigma_j^y )$. By $\sigma^x,\sigma^y$         
and $\sigma^z$ we denote the Pauli matrices. 
Since we restrict ourselves to hermitian boundary terms,
 the values of $\alpha_z$ and $\beta_z$ are real numbers and $\alpha_+$ and  $\beta_+$
are the complex conjugates of $\alpha_-$ respectively $\beta_-$.
In the first publication \cite{paper1} of this series we have shown how to diagonalize $H$
by introducing an auxiliary Hamiltonian
\begin{eqnarray}
\fl
H_{\rm long}=\frac{1}{2}\sum_{j=1}^{L-1}
\left[\sigma_j^+\sigma_{j+1}^-+\sigma_j^-\sigma_{j+1}^+\right] \nonumber \\
+\frac{1}{\sqrt{8}}\left[\alpha_-\sigma_0^x\sigma_1^-+\alpha_+\sigma_0^x\sigma_1^+
+\alpha_z\sigma_1^z+\beta_+\sigma_L^+\sigma_{L+1}^x+\beta_-\sigma_L^-\sigma_{L+1}^x
+\beta_z\sigma_L^z\right]
\label{Hlong}
\end{eqnarray}
(following an idea of \cite{Tj})
which in turn may be diagonalized
in terms of free fermions \cite{JorWig}. 
Note that $H_{\rm long}$ commutes with $\sigma_0^x$ and $\sigma_{L+1}^x$.
Hence the spectrum of $H_{\rm long}$ decomposes into four sectors
$(+,+),(+,-),(-,-),(-,+)$ corresponding to the eigenvalues $\pm 1$ of $\sigma_0^x$
and $\sigma_{L+1}^x$.
The spectrum and the eigenvectors of $H$ are obtained by projecting onto the $(+,+)$--sector.

In this paper we are going to consider the profiles
of the spin operators $\sigma_j^z$ and $\sigma_j^x$.
We have already shown in \cite{paper1} how to compute the
one point function of the $\sigma_j^x$-operator in terms of
pfaffians. The $\sigma_j^z$-profiles are easier to compute.
We will study the ground-state profiles  as well as
 the profiles
of certain excited states of $H$.
The excited states of $H$ we are going to consider 
 correspond to single fermion excitations with respect to the ground-state  
of $H_{\rm long}$, which are characterized by a non-vanishing 
excitation energy as $L$ goes to infinity and which correspond 
to boundary bound states.

We are able to obtain exact results for the ground-state profiles only for
certain values of the boundary parameters \cite{paper1}.
In the case of the $\sigma_j^z$-profiles we will
find exact expressions on the finite chain,
in the continuum limit $L\gg1, z=j/L$  fixed and in the limit $L\to \infty, j\gg 1$.
The 
$\sigma_j^x$-profiles  are given 
in terms of pfaffians of $2j\times 2j$ matrices, that we have 
 not been able to compute  on the finite chain 
or in the continuum limit exactly, therefore our exact results for the $\sigma_j^x$-profiles
 are
restricted to the limit $L\to \infty, j\gg 1$.

A combination of extensive numerical computations 
(for lattice lengths $L\sim 600$ )
and of our  exact results
has led us to conjectures for the profiles of $\sigma_j^z$ and $\sigma_j^x$ in the continuum limit
for more general values of the boundary parameters. Note that once we 
know the $\sigma_j^x$-profile for one specific choice of the boundary terms,
we know already the profile of $\sigma_j^y$ for the boundary terms with $\sigma^x$ and
$\sigma^y$ interchanged in \eref{HXX}.

We will discuss our results for the ground-state profiles 
in the context of conformal field theory.
According to Burkhardt and Xue \cite{BX1,BX2}, in the continuum limit $z=j/L, L\gg 1$ the general form
 of the profile  of any scalar scaling operator $\phi$  
with  bulk-scaling dimension  $x_{\phi}$ 
is given by
\begin{equation}
\langle \phi(j) \rangle_{ab}=[(L/\pi)\sin(\pi z)]^{-x_{\phi}}
    \mathcal{F}_{ab}(z).
\label{genprofil}
\end{equation}
The functional form of $\mathcal{F}_{ab}(z)$ 
depends on the boundary conditions $a$ and $b$ at the left
respectively the right boundary. 
We have shown in \cite{paper2} that  the XX--model with hermitian
boundaries corresponds
to the free compactified boson on a cylinder with  Neumann-Neumann,
Dirichlet-Neumann or Dirichlet-Dirichlet boundary conditions.
The bulk-scaling dimensions $x_{\phi}$ obtained from the two-point functions
for the periodic chain 
are $1$ respectively $\frac{1}{4}$ for $\sigma_j^z$ respectively $\sigma_j^x$ \cite{LSM,McCoy}.

Conformal field theory makes more restrictive predictions
for the behaviour of the profiles of scaling operators near the boundary.
From \eref{genprofil} we obtain that near the boundary the profile is given by
\begin{equation}
\langle\phi(j)\rangle_{ab}=j^{-x_{\phi}}\psi_{ab}(z) \qquad z \ll 1 .
\label{scaling1}
\end{equation}
It was already argued in \cite{FG} that 
if the scaling 
 function $\psi_{ab}(z)$ is different from zero near the boundary, it has the asymptotic form
\begin{equation}
\psi_{ab}(z)=A[1+\mathcal{B}_{ab}^{\phi}z^2+\cdots] \qquad z \ll 1 .
\label{scaling2}
\end{equation}
Furthermore it has been shown in \cite{BX1,BX2,Cardyprofile} that 
 the amplitude 
 $\mathcal{B}_{ab}^{\phi}$ is related to the Casimir amplitude
$\mathcal{A}_{ab}$ via
\begin{equation}
\mathcal{B}_{ab}^{\phi}/\mathcal{A}_{ab}=
-4\pi x_{\phi}/c 
\label{ratioAB}
\end{equation}
where $c$ denotes the central charge of the theory. 
The Casimir amplitude $\mathcal{A}_{ab}$ is defined by the large $L$ behaviour
of the ground-state energy $E_0$, i.e.
\begin{equation}
E_0=e_{\infty}L+f_{ab}+\mathcal{A}_{ab} L^{-1} + \cdots
\end{equation}
where by $e_{\infty}$ and $f_{ab}$ we denote the bulk respectively the
surface free energy. The Casimir amplitudes for the
boundary terms of the Hamiltonian \eref{HXX} can be read off from the partition functions
obtained in \cite{paper2,alcatraz}.

As we are going to show, when dealing with boundaries,
it is not a priori clear which combinations of  spin operators
correspond to the scaling operators of the continuum theory.
Our conjectures for the continuum limit of the profiles will yield,
 that the profiles of $\sigma_j^x$ and 
of $\sigma_j^z$ indeed have the form of \eref{genprofil} (see however equation \eref{connnz}).

We will see that if the boundary conditions we are going to consider correspond 
to Dirichlet-Neumann of the free boson field,
the profiles satisfy \eref{scaling2} and \eref{ratioAB}.
We will also see that for the boundary terms corresponding to
  Dirichlet-Dirichlet and Neumann-Neumann boundary 
conditions the profiles
show a more complicated behaviour near the boundary.
In these cases the equations \eref{scaling2} and \eref{ratioAB} do not hold
in general for the profiles of $\sigma_j^x$ and $\sigma_j^z$.

In the case of Dirichlet-Dirichlet boundary conditions we have 
to build an appropriate linear combination of $\sigma_j^z$ and $\sigma_{j+1}^z$
in order to obtain the correct behaviour of a scaling operator. 
However, in the generic case the linear combination we are going to obtain works only 
at one boundary at a time.
The $\sigma_j^x$-profile vanishes exactly in this case.

For the case of Neumann-Neumann boundary conditions we obtain the 
correct behaviour by considering a linear combination
of $\sigma_j^x$ and $\sigma_j^y$. Again this works only at one boundary at a
time in the generic case.
Here the $\sigma_j^z$-profile vanishes if no diagonal boundary terms 
are present at the boundary (i.e. $\alpha_z=\beta_z=0$).
However, we have also studied a type of boundaries corresponding to
the Neumann-Neumann boundary conditions, where diagonal and non-diagonal
boundaries are present at a boundary at the same time. 
In this case the profile of the $\sigma_j^z$--operator 
vanishes only in the leading order 
and
is determined by a secondary field instead by a primary field.

Our results in the limit $L \to \infty, j\gg 1$
 are in  agreement with the results of Affleck \cite{affleckx}.
He has
considered the semi-infinite XXZ spin-1/2 chain with a transverse magnetic field applied at
the end of the chain (this corresponds to $\alpha_+=\alpha_-\neq 0$,
$\alpha_z=\beta_z=\beta_+=\beta_-=0$ and $L\to \infty$ in our notation).
Affleck has used the bosonization technique, whereas our results have been
 obtained on the lattice.
Assuming Neumann boundary conditions
for the boson field,
he obtained the $\sigma_j^x$-profile for large values of $j$.
At the free fermion point the result  reads 
\begin{equation}
\label{affx}
\langle \sigma_j^x \rangle =  (-1)^j  C j^{-\frac{1}{4}}  
\end{equation} 
where the constant $C$ does not depend on the strength of the field and at the free fermion point
  its numerical value is
given by
\begin{equation} 
\label{C}
C=0.912\,171\,210\,446 \cdots .
\end{equation}
The case of a longitudinal
magnetic field at the end of the chain has also been considered in \cite{affleckx}
(corresponding to $\alpha_z\neq 0$, $\alpha_+=\alpha_-=\beta_z=\beta_+=\beta_-=0$
and $L\to \infty$).
Imposing Dirichlet boundary conditions for the boson field,
the large $j$ behaviour of the $\sigma_j^z$-profile
has been evaluated for small values of $\alpha_z$, yielding
at the free fermion point 
\begin{equation}
\label{affz}
\langle \sigma_j^z \rangle \approx B \alpha_z (-1)^j j^{-1} 
\end{equation}
where the value of $B$ is undetermined. 
The case of diagonal and non-diagonal boundary terms at the boundary at the same time has not 
been considered.
The boundary conditions where we have been able to compute
the profiles exactly 
in the limit $L\to \infty, j\gg 1$ 
can always be mapped onto the boundaries treated by Affleck.
 Our results obtained on the lattice agree with \eref{affx} and \eref{affz}. 
For the $\sigma_j^z$-profiles we have not restricted ourselves to small
values of $\alpha_z$ and we have derived the $\alpha_z$-dependence of $B$.

As already mentioned above, we will also consider the magnetization profiles of
certain eigenstates of $H$, which are obtained by the excitation of a massive fermion
with respect to the  ground-state of $H_{\rm long}$.
The existence of such fermions has already been mentioned in \cite{paper1} 
for special values of the boundary parameters.
Before considering the profiles,
we are going to study the appearance of such fermions
for general, hermitian boundary terms. 

We will see that massive fermions appear if the values of certain functions
of the boundary parameters become greater than a certain threshold.
The respective excitation masses will turn out to depend only on the
boundary parameters of one boundary at a time.
In the general case (non-diagonal together with diagonal boundary terms)
we will find at most one mass per boundary.

In the case 
of purely diagonal boundary terms 
 the fermion masses correspond exactly to the 
non-vanishing energy gaps obtained in the Bethe Ansatz  
 by comparing the energy of the reference state with the energy
of the one-magnon excitations, which correspond to
  boundary 1-strings \cite{SkoSa}.

We will see, that the magnetization profiles
for the excited states under study here,
show 
an exponential fall-off into the bulk for diagonal as well as for non-diagonal 
boundary conditions (the profiles for the non-diagonal boundary terms 
have been obtained numerically).
Hence we are going to refer to these states  as to boundary bound states.
These profiles appear also for $H$ since the excited states we are going 
to study lie in the $(+,+)$-sector of the Hilbert space of $H_{\rm long}$
and do therefore correspond directly to eigenstates of $H$.

This paper is organized as follows:
We start with the exact results in section 2.
These results are restricted to certain choices of boundary terms.
In section 3 we will give our conjectures for the profiles
in the continuum limit for more general
types of boundaries.
We are going to discuss these results in the context of conformal field theory
in section 4.
Boundary bound states are the subject of section 5.
Our conclusions will be given in section 6.
The details of our calculations will be given in the appendix.
In Appendix A we shortly discuss the diagonalization of $H_{\rm long}$ 
and the projection onto the $(+,+)$-sector in order
to provide the basic facts and notations being necessary to follow our
calculations. 
Appendix B deals with the exact computations of 
the $\sigma_j^z$-profiles. How we arrived at our exact results 
for the $\sigma_j^x$-profiles is shown in Appendix C.
In Appendix D we discuss the numerical verification
of our conjectures of the ground-state profiles in the continuum limit.

\section{Exact results}
\label{exfini}
For technical reasons explained in \ref{techsum}, we have been able to obtain analytical results only
 for certain choices of boundary terms (see \tref{paraz} and \tref{parax}).
We will begin with the results for the $\sigma_j^z$-profiles on the finite chain,
 before
turning to the respective expressions in the continuum limit.
Thereafter we will give our exact results  in the limit $L\to \infty, j\gg 1$
for the profiles  of $\sigma_j^x$ and $\sigma_j^z$.

\subsection{Exact results on the finite lattice ($\langle\sigma_j^z\rangle$- profiles)}
\label{exz}
We have computed the exact profiles of $\sigma_j^z$ for values of the boundary parameters given 
in \tref{paraz} \footnote{The cases denoted by a,b,c,d,e 
correspond to the cases $2,4,11,14,16$ in the notation of \cite{paper1,paper2}.}.
Note that for technical reasons (see \ref{calproz} for details) our result for case $e$ 
 is restricted to the choice $\delta=0$ in this section.
Furthermore, observe that the ground-state is twofold degenerate 
for case $d$ with $L$ odd and for case $e$ with $L$ even, due to the existence of additional 
zero modes \cite{paper1}.
In these cases, we have obtained the profiles for the two ground-states, which 
have a well-defined fermion number (see \ref{calproz}).
These profiles are indicated by a $\pm$-sign in the respective expressions.

\begin{table}
\caption{The values of the boundary parameters where we have computed 
the ground-state profiles of $\sigma_j^z$ exactly on the finite chain
and in the continuum limit. By DN (Dirichlet-Neumann) and DD (Dirichlet-Dirichlet) we denote the respective
boundary conditions on the free boson \cite{paper2}. For case e there appears a free, real parameter $\delta$.  
} 
\begin{indented}
\item[]\begin{tabular}{@{}lcclll}
\br
case & $\alpha_z$ & $\beta_z$ & $\beta_+=\beta_-$ & $\alpha_{\pm}$ &  \\
\mr
a & $1\over{\sqrt{2}}$ & 0 & $1\over{\sqrt{2}}$  & 0 & DN \\
b & $1\over{\sqrt{2}}$ & 0 & 1  & 0 & DN \\
c& $1\over{\sqrt{2}}$ & 0 & 0 
    & 0 & DD \\
d& $1\over{\sqrt{2}}$ & $-1\over{\sqrt{2}}$ & 0 
    & 0 & DD  \\
e& $\frac{1}{\sqrt{2}} \tan(\frac{\pi}{4}+\frac{\delta}{2})$ &
   $\frac{1}{\sqrt{2}}\tan(\frac{\pi}{4}-\frac{\delta}{2})$ & 0 
    & 0 & DD \\
\br
\end{tabular}
\end{indented}
\label{paraz}
\end{table}
\begin{itemize}
\item Case a
\begin{equation}
\langle\sigma_j^z\rangle=\frac{1}{2L+2}\left((-1)^L\tan\frac{\pi}{4L+4}+(-1)^j\cot\frac{(2j-1)\pi}{4L+4}\right).
\label{exa}
\end{equation}
\item Case b 
\begin{equation}
\langle\sigma_j^z\rangle=\frac{(-1)^j}{2L+1}\cot\frac{(2j-1)\pi}{4L+2}.
\label{exb}
\end{equation}
\item Case c
\begin{equation}
\langle\sigma_j^z\rangle=\frac{1}{2L+1}\left((-1)^L+(-1)^j\left(\sin\frac{(2j-1)\pi}{4L+2}\right)^{-1}\right).
\label{exc}
\end{equation}
\item Case d, 
$L$ odd: The ground-state is twofold degenerate in this case.
\begin{equation}
\langle\sigma_j^z\rangle_{\pm}=\frac{1}{L}\left((-1)^j\cot\frac{(2j-1)\pi}{2L} 
                                           \pm 1\right).       
\label{exdodd}
\end{equation}
\item Case d, $L$ even 
\begin{equation} 
\langle\sigma_j^z\rangle=\frac{(-1)^j}{L}\left(\sin\frac{(2j-1)\pi}{2L}\right)^{-1}.
\label{exdeven}
\end{equation}
\item Case e, 
$L$ odd, $\delta=0$ : The same as given for case d, $L$ even. 
\item Case e,
$L$ even, $\delta=0$ : The same as given for case d, $L$ odd.
 The ground-state is twofold degenerate in this case, too.
\end{itemize}

\subsection{Exact results in the continuum limit}
The continuum limit defined by $L\gg 1, z=j/L$ fixed
 of the expressions  \eref{exa}-\eref{exdeven} is readily 
obtained. Furthermore  
we have been able to compute the profile 
in the continuum limit for case e for general values of $\delta$ (see \ref{calproz}).

\begin{itemize}
\item Case a and case b 
\begin{equation}
\langle\sigma_j^z\rangle=(-1)^j\left(L\sin(\pi z)\right)^{-1}
\cos^2\left(\frac{\pi z}{2}\right)+\Or(1/L^2).
\label{continuumz2+4}
\end{equation}

\item Case c
\begin{equation}
\langle\sigma_j^z\rangle=(-1)^j\left(L\sin(\pi z)\right)^{-1}
\cos\left(\frac{\pi z}{2}\right)+ (-1)^L/2L+ \Or(1/L^2).
\label{continuumz11}
\end{equation}

\item Case d: The same expression as given for case e with $\delta=0$ and $L$ even
and $L$ odd interchanged.

\item Case e, $L$ odd
\begin{equation}
\langle \sigma_j^z \rangle =(-1)^j \cos \delta
\left(L\sin(\pi z)\right)^{-1}  + \Or(1/L^2).
\label{continuumz16odd}
\end{equation}

\item Case e, $L$ even
\begin{equation}
\langle \sigma_j^z \rangle_{\pm} =
 (-1)^j \cos(\delta\mp\pi z) \left(L\sin(\pi z)\right)^{-1}
\pm 1/L + \Or(1/L^2).
\label{sigmaz_case16_staggerd_Leven}
\end{equation}

\end{itemize}

\subsection{Exact results in the limit $L\to \infty, j\gg 1$}

We have obtained exact results for the profiles of $\sigma_j^z$
in the limit $L\to \infty, j\gg 1$ for the five cases given in \tref{paraz}.
For the cases a,b,c,d the result may be read of from the expressions 
obtained on the finite lattice.  
This can not be done for case e with arbitrary values of $\delta$,
since the respective result on the finite lattice is restricted to $\delta=0$.
However, in the limit $L\to \infty, j\gg 1$ the calculation for case e
can be done for arbitrary values of $\delta$ (see \ref{calproz}). 
Using $\alpha_z$ instead of the parameter $\delta$ (see \tref{paraz} 
for the relation between $\delta$ and $\alpha_z$) we have obtained
\begin{equation}
\label{zt}
\langle \sigma_j^z \rangle = (-1)^j 
\frac{\sqrt{8}\alpha_z}{(1+2\alpha_z^2)\pi}j^{-1}  + \cdots .
\end{equation}
The profiles for the cases a,b,c,d are also given by this expression if
one replaces $\alpha_z$ by $\frac{1}{\sqrt{2}}$.
The result \eref{zt} has to be compared to \eref{affz}, which yields $B$ 
as a function of $\alpha_z$,  i.e.
\begin{equation}
B=
\frac{\sqrt{8}}{(1+2\alpha_z^2)\pi} .
\end{equation}
\begin{table}
\caption{Boundary parameters for the cases where we 
found exact expressions for the ground-state profiles of $\sigma_j^x$
in the limit $L\to \infty, j\gg 1$.
The first case corresponds to   Dirichlet-Neumann (DN)
boundary conditions, whereas the second case corresponds
to Neumann-Neumann (NN) boundary conditions on the free boson.
The free parameters $\chi$ and $\varphi$ are real constants.
}
\begin{indented}
\item[]
\begin{tabular}{@{}lllllll}
\br
case &  $\alpha_+$ & $\alpha_-$ & $\beta_+$ & $\beta_-$ &   \\
\mr
f &  1 & 1 & 0 & 0 & DN \\
g &  $\rme^{\rmi \varphi}$ & $\rme^{-\rmi \varphi}$ & 
      $\rme^{\rmi(\chi+\varphi)}$ & $\rme^{-\rmi(\chi+\varphi)}$ & NN \\
\br
\end{tabular}
\end{indented}
\label{parax}
\end{table}
We have obtained exact results for
 the profiles of $\sigma_j^x$  in the limit
$L\to \infty$ and $j \gg 1$
 for the two
cases given in \tref{parax}\footnote{In the notation of \cite{paper1,paper2} 
these cases correspond to case 2 respectively case 9.}. 
Our computations for case f (see \ref{casef}) yield
\begin{equation}
\langle \sigma^x_j \rangle = (-1)^j  A j^{-\frac{1}{4}} +\cdots
\label{c2t}
\end{equation}
where $A$ is given by
\begin{equation}
 A= \rme^{1/4}2^{7/12}C_G^{-3}
\label{lnA}
\end{equation}
and $C_{\rm G}$ is Glaishers constant \cite{Gla}, which can be approximately given by
\begin{equation}
\label{G}
C_{\rm G}= 1.282\, 427 \, 129 \, 100 \, 62 \cdots .
\end{equation} 
For case g we have obtained (see \ref{caseg})
\begin{equation}
\label{cgt}
\langle \sigma^x_j \rangle = (-1)^j  A \cos \varphi j^{-\frac{1}{4}} +\cdots .
\end{equation}
Comparing the value of $C$ in \eref{C} 
and the value of $A$ in \eref{lnA} using \eref{G} shows  that our results \eref{c2t} and
\eref{cgt} for $\varphi=0$ are in agreement with \eref{affx}.

At this point we want to draw the readers attention 
to an interesting by-product we have obtained during 
our computations for case g. 
It is of major interest for our conjecture
of the continuum limit of the $\sigma_j^x$ profiles in the case of Neumann-Neumann
boundary conditions presented in the next section.
We have obtained that the ground-state profile of $\sigma_j^x$ for arbitrary 
values of $\varphi$ and $\chi$ (see \tref{parax}) is given by
\begin{equation}
\langle \sigma_j^x  \rangle=
\cos\left(\varphi+\frac{\chi+2m\pi}{L+1}j\right) f(j,L)
\label{sx9}
\end{equation}
where $f(j,L)$ does not depend on $\varphi,\chi$ and may be computed numerically
(see \ref{caseg}).
The value of $m$ depends on the value of $\chi$ and on the lattice length $L$.
For odd $L$ we have $m=0$ for $-\pi<\chi<\pi$.
If $\chi=\pi$ the ground-state of $H$ is twofold degenerate.
The value of $m$ is either $m=0$ or $m=-1$.
For even $L$ the value of $m$ is $\frac{1}{2}$ for $0>\chi>-\pi$ and $-\frac{1}{2}$
for $0<\chi\leq \pi$.
Here the ground-state is twofold degenerate for $\chi=0$ the value
of $m$ being either $\frac{1}{2}$ or $-\frac{1}{2}$.

\section{Conjectures}
Our results given so far allow to conjecture  the ground-state profiles in the continuum limit
for more general values of the boundary parameters. 
We have checked our results by computing the 
profiles on the finite chain numerically 
 for lattice lengths up to $L=700$.
Thereafter we have used the finite size data to extrapolate 
the profiles for  $L\to\infty$ and fixed values of $z=j/L$.
We have performed these extrapolations for 
 $150$ different values of the boundary parameters
each with $19$ (L even) respectively $20$ (L odd)  values of $z$.
The accuracy of our extrapolations varies with 
the choice of the boundary parameters.  
In most of the cases the relative deviation from our conjectures presented in the
following has been
of the order of $10^{-12}-10^{-7}$ (see \ref{cnum} for details).

We have considered the following types of boundary terms:
\begin{itemize}
\item[$\circ$]
Dirichlet-Dirichlet: 
\begin{equation}
\label{dd}
\fl
\alpha_{\pm}=\beta_{\pm}=0 \qquad \alpha_z=\frac{1}{\sqrt{2}}\tan\left(\frac{\omega+\delta}{2}\right)
\qquad \beta_z=\frac{1}{\sqrt{2}}\tan\left(\frac{\omega-\delta}{2}\right)
\end{equation}
Note that for $\omega=\frac{\pi}{2}$ we recover the boundary terms of case e in \tref{paraz}.
\item[$\circ$]
Dirichlet-Neumann: 
Here we have considered only a special type of boundary terms, i.e.
\begin{equation}
\label{dn1}
\beta_+=\beta_-\neq 0 \qquad \beta_z=\alpha_{\pm}=0 \qquad \alpha_z \mbox{ free}
\end{equation}
\item[$\circ$]
Neumann-Neumann: We have studied two types of boundary terms, i.e.
\begin{equation}
\label{nn1}
\alpha_z=\beta_z=0 \qquad \alpha_{\pm}=R_{\alpha}\rme^{\pm\rmi\varphi}\qquad
                        \beta_{\pm}=R_{\beta}\rme^{\pm\rmi(\chi+\varphi)}
\end{equation}
respectively
\begin{equation}
\alpha_z=-\beta_z\qquad \alpha_+=\alpha_-=\beta_+=\beta_-\neq 0 .
\label{nn2}
\end{equation}
\end{itemize}
Note that the boundary terms  \eref{nn1} are identical to the boundaries
of case g in \tref{parax} if one sets $R_{\alpha}=R_{\beta}=1$, whereas
  no case in \tref{paraz} and \tref{parax} corresponds
to boundary terms which satisfy \eref{nn2}.  
In this case our result is based on purely numerical computations.
Observe also, that this is the only type of boundary terms with diagonal
and non-diagonal terms at  one end of the chain, which is studied in this paper.

\subsection{Dirichlet-Dirichlet boundary conditions}
For this type of boundary conditions the ground-state profile of $\sigma_j^x$ 
vanishes if the ground-state is non-degenerate, since the Hamiltonian
commutes with $\sigma_1^z\sigma_2^z\cdots \sigma_L^z$.
Therefore we have restricted ourselves to the ground-state profiles of $\sigma_j^z$.
Our results for the cases c,d,e from \tref{paraz} and our numerical investigations 
suggest 
 the $\sigma_j^z$-profiles for general values of $\alpha_z$ and $\beta_z$, i.e. 
\begin{equation}
\langle \sigma_j^z \rangle =(-1)^j \frac{\sin(\omega+\delta-2(\omega+Q\pi)z)}{L\sin(\pi z)} 
+\frac{2(\omega+Q\pi)}{\pi L} .
\label{conconz}
\end{equation}
This equation is valid for $L$ even and $L$ odd.
 The value of $Q$ depends on the values of $\omega$ 
and the lattice length $L$.
 This dependence is given in \tref{omegaprime}.
The discontinuities are due to level crossings in 
the spectrum at the respective values of $\omega$.
At these points ($\omega=0$ for $L$ odd and $\omega=\pm \pi/2$
for $L$ even) the ground-state is twofold degenerate.

\begin{table}
\caption{Dependence of $Q$ on $\omega$ and $L$.}
\begin{indented}
\item[]\begin{tabular}{@{}lcccc}
\br
 & $-\pi<\omega<-\frac{\pi}{2}$ & $-\frac{\pi}{2}<\omega<0$
 & $ 0<\omega<\frac{\pi}{2}$ & $\frac{\pi}{2}<\omega<\pi$ \\
\mr
$L$ even & $Q=1$ & $Q=0$ & $Q=0$ & $Q=-1$ \\
$L$ odd  & $Q=\frac{1}{2}$ & $Q=\frac{1}{2}$ &
 $Q=-\frac{1}{2}$ & $Q=-\frac{1}{2}$ \\
\br
\end{tabular}
\end{indented}
\label{omegaprime}
\end{table}

\subsection{Dirichlet-Neumann boundary conditions}
Our numerical computations suggest the following profile of the
$\sigma_j^x$-operator:
\begin{equation}
\label{condnx}
\langle \sigma_j^x \rangle =
\pi^{1/4}A(-1)^{L+1-j}\left[\sin\left(\frac{\pi z}{2}\right)\right]^{1\over{2}}
\left( L \sin ( \pi z )\right)^{-\frac{1}{4}}
\end{equation}
where $A$ is given by \eref{lnA}.  

 For two cases corresponding to this type of boundaries we have already 
considered the exact $\sigma_j^z$-profiles, namely
for case a  and b  from \tref{paraz}. 
The given expression \eref{continuumz2+4} suggests universal behaviour with respect to the variation of the value of
$\beta_{\pm}$. Our numerical calculations 
for various values of $\beta_{\pm}$  confirmed our guess. Furthermore 
they  yield the dependence of the profile on the second parameter
$\alpha_z$. We have obtained 
\begin{equation}
\label{condnz}
\langle\sigma_j^z\rangle=(-1)^j \frac{\sqrt{8} \alpha_z}{1+2\alpha_z^2}
\cos^2\left(\frac{\pi z}{2}\right) \left(L\sin(\pi z)\right)^{-1} .
\end{equation}
Note that the form of the profile does not depend 
on the value of $\alpha_z$. However, observe also that the 
amplitude does.
For $\alpha_z=0$ the profile vanishes, since then 
the Hamiltonian $H$
commutes with $\sigma_1^x\sigma_2^x\cdots\sigma_L^x$ (the ground-state of $H$ is
non-degenerate \cite{paper1}).

\subsection{Neumann-Neumann boundary conditions}
Our exact result for case g from \tref{parax}, 
 in particular equation \eref{sx9}, in combination with numerical 
computations yields the $\sigma_j^x$-profile for the first type 
of boundaries in question (see \eref{nn1}):
\begin{equation}
\label{connnx}
\langle  \sigma_j^x  \rangle= \pi^{1/4}A(-1)^j
\cos\left(\varphi+(\chi+2m\pi)z \right)  \left(L\sin(\pi z)\right)^{-{1\over{4}}}
\end{equation}
where the value of $m$ depends on the values of $\chi$ and the lattice length $L$ 
as already discussed at the end of section 2.3.
The $\sigma_j^z$ profiles vanish, as long as the ground-state
is non-degenerate (see \ref{calproz} for details). 
We did not consider the $\sigma_j^z$ profiles where the ground-state 
is degenerate.

For the second type of boundary terms  \eref{nn2} the $\sigma_j^z$ profile does not
vanish. 
For odd values of $L$ our numerical investigations  suggest
\begin{equation}
\label{connnz}
\langle \sigma_j^z \rangle 
=(-1)^j\pi\sqrt{2}\frac{\alpha_z}{\alpha_+^2}\cos(\pi z) (L \sin (\pi z))^{-2}  .
\end{equation}
We have not been able to find an analytic expression for the profile 
for even values of $L$.
For $\alpha_+=0$ the prefactor in \eref{connnz} diverges. However, in this case the
profile is already given by \eref{conconz}. 
For $\alpha_z=0$ the profile vanishes as in the case of the Dirichlet-Neumann boundary
condition \eref{dn1}, since in this case 
the Hamiltonian $H$ has a higher symmetry. It
commutes with $\sigma_1^x\sigma_2^x\cdots\sigma_L^x$
 (the ground-state of $H$ is also
non-degenerate here \cite{paper1}).

 Comparing \eref{connnz} with \eref{genprofil}  
yields that the leading term of the $\sigma_j^z$-profile is zero as for the boundary terms \eref{nn1},
since according to the long range behaviour of the two-point function
 the bulk-scaling dimension $x_{\phi}$ associated to the $\sigma_j^z$ operator is $1$.
Since there  exists no spinless primary field with bulk-scaling dimension $2$ \cite{AlBaGrRi} 
we conclude that the profile \eref{connnz} is determined by a secondary field 
(the  profile of a primary field with spin
vanishes \cite{BX2}).

We have checked the $\sigma_j^x$-profile only for odd $L$.
It is not affected by the presence
of the diagonal terms in \eref{nn2} and is already given by \eref{connnx} using $\varphi=\chi=m=0$.

\section{Comparison with the predictions of Conformal Field Theory}
In this section we are going to consider the behaviour of the profiles 
we have conjectured near the boundaries, i.e. we consider the limit $z\ll 1$
 (respectively $(1-z)\ll 1$) , in order to check for the 
validity of the equations \eref{scaling2} and \eref{ratioAB}.

\subsection{Dirichlet-Dirichlet boundary conditions}
The behaviour of \eref{conconz}  near the left boundary is readily obtained:
\begin{equation}
\fl
\langle \sigma_j^z \rangle = (-1)^j \frac{\sin(\omega+\delta)}{j\pi}
 \left(1-\cot(\omega+\delta)2\omega' z
+\left(\frac{1}{6}-\frac{2\omega'^2}{\pi^2}\right)(\pi z)^2 +\cdots \right) +\frac{2\omega'}{\pi L}.
\label{boundaryz_conjecture}
\end{equation}
where $\omega'=\omega+Q\pi$.
The behaviour of the profile near the right boundary may be obtained by exchanging $\delta$ by
$-\delta$ and performing the transformation $j \mapsto L+1-j$.
This can be seen from \eref{dd}.
Note that \eref{boundaryz_conjecture} does not agree with equation \eref{scaling2}, since in \eref{scaling2}
there appears no term which is linear in $z$.
This problem is avoided by considering the profile of the following linear combination of
$\sigma_j^z$ and $\sigma_{j+1}^z$ for odd values of $j$:
\begin{equation}
\fl
\left\langle \tan\left(\frac{\omega+\delta}{2}\right)\sigma_j^z-
\cot\left(\frac{\omega+\delta}{2}\right)\sigma_{j+1}^z\right\rangle =
-\frac{2}{j\pi}\left(1+ \left( \frac{1}{6} - \frac{2\omega'^2}{\pi^2}\right)(\pi z)^2 +\cdots \right).
\label{combizprofil}
\end{equation}
The value of the Casimir amplitude $\mathcal{A}_{ab}$ can be obtained from
the results in \cite{paper2,alcatraz}.
It is given by $\mathcal{A}_{ab}=-\frac{\pi}{24}+\frac{\omega'^2}{2\pi^2}$.
Hence \eref{combizprofil}
 is in agreement with \eref{ratioAB}.
Note that this
operator does not yield 
the correct scaling behaviour near the right boundary, since the given linear 
combination of $\sigma_j^z$ and $\sigma_{j+1}^z$  depends on the value of $\delta$,
which enters the left boundary term in a different way than the right boundary term (see \eref{dd}).
Another linear combination can be chosen to obtain the correct behaviour near the right boundary.

\subsection{Dirichlet-Neumann boundary conditions}
The  Casimir amplitude for this type of boundary 
conditions is independent from the explicit values 
of the boundary parameters.
Its value is given by $\mathcal{A}_{ab}=\frac{\pi}{48}$ \cite{paper2}.
The profiles \eref{condnx} and \eref{condnz}  vanish near the left respectively near
the right boundary  
and therefore the predictions  \eref{scaling2} and \eref{ratioAB} are not valid.
Near the right boundary we obtain  from \eref{condnx} 
\begin{equation}
 \langle   \sigma_{L+1-j}^x  \rangle= (-1)^j
A   j^{-\frac{1}{4}} \left( 1-\frac{(\pi z)^2}{48} +\cdots \right)
\end{equation}
whereas \eref{condnz} yields near the left boundary
\begin{equation}
\langle \sigma_j^z \rangle =
(-1)^j \frac{\sqrt{8} \alpha_z}{1+2\alpha_z^2} ( j\pi)^{-1} \left( 1- \frac{(\pi z)^2}{12}+\cdots \right) .
\label{boundaryz241}
\end{equation}
Both expressions are in agreement with the predictions \eref{scaling2} and \eref{ratioAB}.

\subsection{Neumann-Neumann boundary conditions}
From \eref{connnx} we obtain the profiles 
of $\sigma_j^x$ near 
the left boundary, i.e.
\begin{eqnarray}
\label{combixprofil}
\fl
\langle \sigma_j^x  \rangle=(-1)^j
A   j^{-\frac{1}{4}}\cos\varphi \nonumber \\
\times\left( 1 - \tan(\varphi)(\chi+2m\pi)z
+\left(\frac{\pi^2}{24}-\frac{(\chi+2m\pi)^2}{2}\right)z^2 + \cdots \right).
\end{eqnarray}
The Casimir amplitude has been shown in \cite{paper2} to be
$\mathcal{A}_{ab}=-\frac{\pi}{24}+2\pi\left(m+\frac{\chi}{2\pi}\right)^2$.
Note that
the profile is in disagreement with the predicted behaviour 
for a true scaling operator given in \eref{scaling2}.
In order to obtain the correct behaviour near the boundary,
we have to consider an appropriate linear combination of $\sigma_j^x$ and $\sigma_j^y$.
At the left boundary we obtain
\begin{eqnarray}
\fl
\langle  \cos\varphi \sigma_j^x - \sin\varphi \sigma_j^y \rangle=(-1)^j
A  j^{-\frac{1}{4}}
\left( 1
+\left(\frac{\pi^2}{24}-\frac{(\chi+2m\pi)^2}{2}\right)z^2 + \cdots \right)
\end{eqnarray}
which is in agreement with \eref{scaling2} and \eref{ratioAB}. Note that building
the given linear combination is equivalent to performing a 
rotation around the z-axis such that $\varphi=0$ in \eref{nn1}.
However, similar to the case of the $\sigma_j^z$ profile for the Dirichlet-Dirichlet boundary condition,
this linear combination does not lead to the correct behaviour near the right boundary.
To obtain the correct behaviour near the right boundary one has to perform 
a different rotation as long as $\chi\neq 0$.

It is interesting to consider the respective profile of the
 orthogonal linear combination of $\sigma_j^x$ and $\sigma_j^y$, which yields
\begin{eqnarray}
\langle \sin\varphi \sigma_j^x + \cos\varphi \sigma_j^y  \rangle=-(-1)^j
A   j^{-\frac{1}{4}}
\left( (\chi+2m\pi)z
 + \cdots \right).
\end{eqnarray}
Note that this profile vanishes near the left boundary and therefore 
the predictions \eref{scaling2} and \eref{ratioAB} do not apply.

 


\section{Boundary bound states}
As already mentioned in the introduction, we have also studied the profiles 
of certain excited states of $H$. These 
states are obtained by the excitation of a massive fermion 
with respect to the ground-state of $H_{\rm long}$.
We refer to these excitations as to boundary excitations, since
we are going to see, that they give rise to a
 contribution to the profile  which decays exponentially into the bulk.

First, we are going to look for massive  excitations 
for general, hermitian boundary terms in section 5.1 by studying
the secular equation $p(x^2)=0$ which determines the fermion energies and
which has been obtained in \cite{paper1}.
In section 5.2 we will study the magnetization profiles near the boundaries
of the states, where one massive fermion is excited, for two special types
of boundary terms.

In the case of diagonal boundary terms the energy gaps defined with respect to the ground-state
 energy are in one to one correspondence 
to the boundary 1-strings found as solutions of the Bethe Ansatz equations.
The 1-strings define the energy gaps between the energy of
the reference state and the eigenstates  in the one magnon sector.
At the free fermion point 
a boundary 1-string appears if $2\alpha_z^2 >1$ or $2\beta_z^2>1$ \cite{SkoSa}. 
The energy gaps  corresponding to the boundary 1-strings in the limit 
$L\to \infty$ are given by 
\begin{equation}
\label{magmass}
E_{\alpha}=\frac{1}{2}\left(\frac{1}{\sqrt{2}\alpha_z} +\sqrt{2}\alpha_z\right) 
\qquad 
E_{\beta}=\frac{1}{2}\left(\frac{1}{\sqrt{2}\beta_z} +\sqrt{2}\beta_z\right) 
\end{equation}
where by $E_{\alpha}$ and $E_{\beta}$ we denote the energy gaps 
corresponding to the boundary 1-string which appears for $2\alpha_z^2 >1$ respectively
$2\beta_z^2>1$.

\subsection{Massive excitations}
The fermion energies are obtained by
\begin{equation}
\label{fe}
2\Lambda=\frac{1}{2}(x+1/x)
\end{equation}
where the values of $x$ are given by the roots of a polynomial $p(x^2)$ \cite{paper1}.
Since the energies have to be real numbers, the roots of the polynomial lie
either on the unit circle or on the real axis.
We will focus on the roots which lie on the real axis for $L\to \infty$.
Due to \eref{fe} these roots yield fermions with a finite energy in this limit.
We have not been  able to determine whether there
exist massive fermions corresponding to roots lying on the unit circle.

We are going to
restrict ourselves to the search for zeros satisfying $|x|<1$, since 
if $x$ is a zero of $p(x^2)$ so is $1/x$.
In the limit $L\to \infty$, the secular equation
$p(x^2)=0$  reduces to
\begin{equation}
\fl
\left(1+(1-2\alpha_z^2-2\alpha_-\alpha_+)x^2-2\alpha_z^2 x^4\right)
\left(1+(1-2\beta_z^2-2\beta_-\beta_+)x^2-2\beta_z^2 x^4\right)=0 .
\label{polbound}
\end{equation}
Solving for $x^2$ yields two solutions with $|x^2|<1$, i.e.
\begin{equation}
x^2=\frac{1}{4\alpha_z^2}\left[ 1-2\alpha_-\alpha_+ -2\alpha_z^2 +
 \sqrt{ 8\alpha_z^2+(2\alpha_z^2+2\alpha_-\alpha_+ -1)^2} \right]
\label{firstbound}
\end{equation}
and
\begin{equation}
x^2=\frac{1}{4\beta_z^2}\left[ 1-2\beta_-\beta_+ -2\beta_z^2 + 
 \sqrt{ 8\beta_z^2+(2\beta_z^2+2\beta_-\beta_+ -1)^2} \right] .
\label{secondbound}
\end{equation}
Note that these solutions depend only on the parameters of one boundary
at a time.  
The fermion energies are obtained by taking the positive square root 
of \eref{firstbound} respectively \eref{secondbound} and using \eref{fe}.
Hence we conclude that the maximum number of massive excitations obtained
by our ansatz is two (one per boundary). 
 This result is interesting since it confirms an 
analogous result obtained by Ameduri et al \cite{AmKoLe} in their study of the 
boundary sine-Gordon theory at the free fermion point. 

If the value of $\alpha_z$ or $\beta_z$ equals zero, the  expression
\eref{firstbound} respectively \eref{secondbound} has to be replaced 
by
\begin{equation}
\label{bd}
x^2=1/(2\alpha_-\alpha_+ -1)
\quad
\mbox{respectively}\quad
x^2=1/(2\beta_-\beta_+ -1) .
\end{equation}
If $\alpha_-=0$ or $\alpha_+=0$ respectively $\beta_-=0$ or $\beta_+=0$ 
the  result simplifies to 
\begin{equation}
\label{bn}
x^2=1/(2\alpha_z^2) \quad \mbox{respectively} \quad
x^2=1/(2\beta_z^2) .
\label{boundz}
\end{equation}
All solutions of \eref{polbound} are only asymptotic solutions of the
original polynomial if they satisfy the assumption $|x|<1$.
In the case of diagonal boundary terms this condition coincides with the 
condition for the existence of boundary 1-strings in the Bethe Ansatz.
The energies of the one-magnon excitations corresponding to the boundary 1-strings
given in \eref{magmass}
correspond exactly to the fermion energies \eref{fe} obtained by the roots in \eref{boundz}.

\subsection{Magnetization profiles near the boundary}
We have studied the magnetization profiles for certain states, which differ from the ground-state
of $H_{\rm long}$ by a boundary excitation,
for two types of boundaries. Note that the ground-state of $H_{\rm long}$ is at least twofold
degenerate. We have chosen it such that the excited state in question lies in the
 $(+,+)$-sector. Therefore these states correspond also to eigenstates
of $H$.  
First we have considered diagonal boundary terms 
\begin{equation}
\label{bz}
\alpha_z=1/(2\beta_z)=\rme^{\xi}/\sqrt{2}\qquad \alpha_{\pm}=\beta_{\pm}=0 
\end{equation}
for even and odd values of $L$.
Without loss of generality we have restricted ourselves to $\xi>0$.
In this case $2\alpha_z^2 $ becomes larger than 1 and according to
\eref{boundz} and \eref{fe}  there exists 
 one boundary excitation with 
energy $2\Lambda=\cosh \xi$.
We are able to compute 
the $\sigma_j^z$-profile analytically
in the limit $L\to \infty, j\gg 1$ (see \ref{calproz}), i.e.
\begin{equation}
\label{bzl}
\langle \xi \mid \sigma_j^z \mid\xi \rangle\approx \frac{(-1)^j}{\pi}
\frac{1}{\cosh\xi} j^{-1}+2(\rme^{2\xi}-1)\rme^{-2j\xi} 
\end{equation}
where $|\xi\rangle$ denotes the state obtained by the excitation of the massive fermion.
Observe that the correlation length $1/(2\xi)$ is related to the excitation 
mass $2\Lambda=\cosh \xi$.
The profile near the right boundary is
\begin{equation}
\label{bzr}
\langle \xi \mid \sigma_{L+1-j}^z \mid\xi \rangle\approx \frac{(-1)^j}{\pi}
\frac{1}{\cosh\xi} j^{-1} 
\end{equation}
as expected.
The $U(1)$ charge, which is given by
the projection of the total spin onto the z-axis, of the state $\mid\xi\rangle$
is $0$ for $L$ even and $\frac{1}{2}$ for $L$ odd.

Notice that the 
 exponential part in \eref{bzl} is not present if one considers
the respective ground-state profile (see \eref{zt}).
We conclude, that adding a massive excitation to the ground-state
yields an additive contribution to the profile which decays exponentially into the bulk.

The second type of boundaries we have considered are non-diagonal and  given by
\begin{equation}
\label{bx}
\alpha_{\pm}=\sqrt{\cosh\xi_{\alpha}\exp\xi_{\alpha}} \qquad
\beta_{\pm}=\sqrt{\cosh\xi_{\beta}\exp\xi_{\beta}}\qquad \alpha_z=\beta_z=0 .
\end{equation}
We restricted ourselves to even values of $L$, since in this case for odd values of $L$ 
the excitation  of a  fermion with respect to the ground-state of 
$H_{\rm long}$ does not lead to a state which lies in the $(+,+)$-sector and
hence does not correspond to an eigenstate of $H$ (see \ref{fuck} for details). 
Since the $\sigma_j^z$-profile vanishes exactly for the boundaries \eref{bx}, we have studied
the $\sigma_j^x$-profile in this case.  In contrast to the $\sigma_j^z$-profiles given
in \eref{bzl} and \eref{bzr} our results for the $\sigma_j^x$-profiles we are going
to present are 
based on numerical computations for $L=\infty$ in the special case $\xi_{\beta}=-\xi_{\alpha}$
and for $L=800$ in the general case.

In \cite{paper1} we have shown that 
the $\sigma_j^x$-profiles for the boundaries 
given in \eref{bx} may be computed in terms of determinants of $j\times j$ matrices.
In the case $\xi_{\beta}=-\xi_{\alpha}$ it is possible to compute these
matrices analytically in the limit $L\to\infty$ (see \ref{fuck}).
Thus  the numerical study of the profile near the boundaries
is easier than in the general case.
Here one boundary excitation appears with energy $2\Lambda=\cosh\xi_{\alpha}$ (see \eref{bd} and \eref{fe}).
For $\xi_{\alpha}>0$ we have obtained by numerical extrapolation from data for $j=1\ldots 100$  
\begin{equation}
\label{bxl1}
\langle \xi_{\alpha} \mid \sigma_j^x \mid \xi_{\alpha} \rangle \approx (-1)^{j+1}
Aj^{-\frac{1}{4}}\left(1-2\rme^{(-1)^j\xi_{\alpha}}\rme^{-2j\xi_{\alpha}}\right)
\end{equation}
and
\begin{equation}
\label{bxr1}
\langle \xi_{\alpha} \mid \sigma_{L+1-j}^x \mid \xi_{\alpha} \rangle \approx (-1)^j
Aj^{-\frac{1}{4}} 
\end{equation}
where $A$ is given by \eref{lnA} and 
$\mid \xi_{\alpha}\rangle$ denotes the state obtained by the excitation 
of the fermion with energy $2\Lambda=\cosh\xi_{\alpha}$. 
Again the exponential term is absent in the respective expressions for the ground-state profile.
Note again the relation between the correlation length and the excitation mass.

Inspired by this result we have also studied the case $\xi_{\alpha}\neq -\xi_{\beta}$.
If $\xi_{\alpha},\xi_{\beta}>0$ there exist two boundary excitations with energies
 $2\Lambda_{\alpha}=\cosh\xi_{\alpha}$ and $2\Lambda_{\beta}=\cosh\xi_{\beta}$ in
the limit $L\to \infty$.
We have considered  the respective magnetization profiles on the finite chain numerically
($L=800$). Exemplarily the figures 1 and 2 show the profiles we have  obtained numerically
 for the state $\mid \xi_{\alpha} \rangle$ 
near the left and the right boundary, where we have chosen $\xi_{\alpha}=0.075, \xi_{\beta}=0.05$.
\begin{figure}[tbp]
\setlength{\unitlength}{1mm}
\def\setl{ \setlength\epsfxsize{8.0cm}}
\begin{picture}(155,90)
\put(28,80){
        \makebox{
                \setl
                \epsfbox{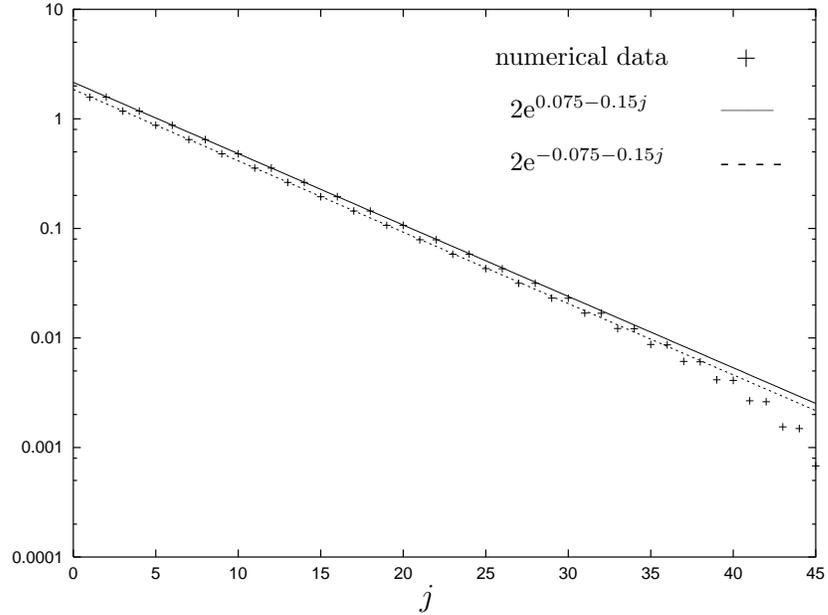}}
        }
\put(10,81){\footnotesize $(-1)^j j^{1/4}A^{-1}\langle\xi_{\alpha}\mid \sigma_j^x \mid \xi_{\alpha}\rangle+1\qquad$}
\put(100,70){\footnotesize numerical data }
\put(132,70){\footnotesize +} 
\put(102,63){\footnotesize $2\rme^{0.075-0.15j}$}
\put(130,63){\footnotesize ------} 
\put(102,56){\footnotesize $2\rme^{-0.075-0.15j}$}
\put(130,56){\footnotesize  - - - - }    
\put(90,-2){$j$}
\end{picture}
\caption{The exponential part of the $\sigma_j^x$--profile of the state $\mid \xi_{\alpha}\rangle$ near  the left boundary 
 for the choice  $\xi_{\alpha}=0.075, \xi_{\beta}=0.05$ in \eref{bx}.
The numerical data (+) have been obtained for $L=800$.  The two lines show the 
behaviour according to \eref{bxl1} for even respectively odd values of $j$.
The deviation of the numerical data from the straight lines for 
large values of $j$ is an
effect of the finite lattice.
        }
\label{left}
\end{figure}
\begin{figure}[tbp]
\setlength{\unitlength}{1mm}
\def\setl{ \setlength\epsfxsize{8.0cm}}
\begin{picture}(155,90)
\put(28,80){
        \makebox{
                \setl
                \epsfbox{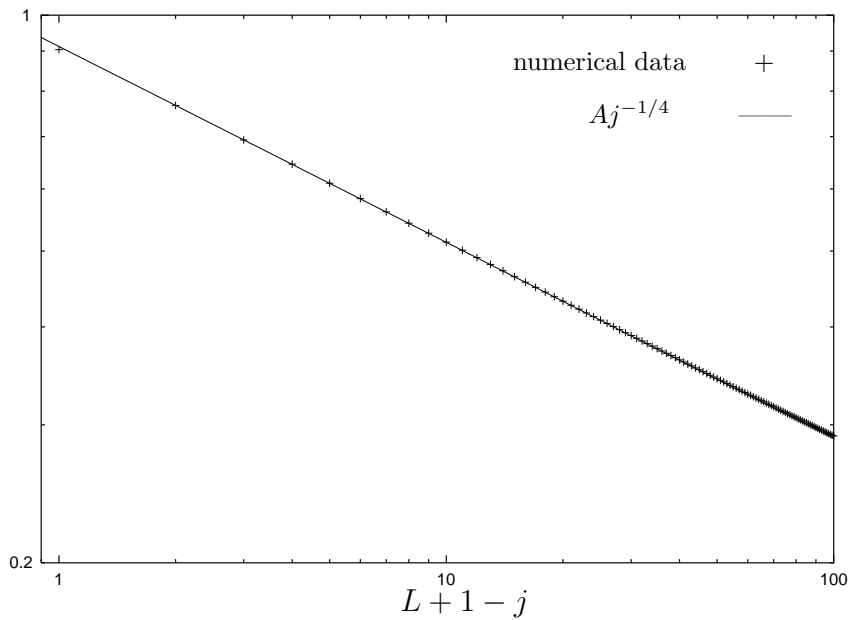}}
        }
\put(10,81){\footnotesize $(-1)^j \langle\xi_{\alpha}\mid \sigma_{L+1-j}^x \mid \xi_{\alpha}\rangle$ }
\put(100,70){\footnotesize numerical data }      
\put(132,70){\footnotesize +} 
\put(110,63){\footnotesize $A j^{-1/4}$}
\put(130,63){\footnotesize ------}
\put(85,-2){$L+1-j$}
\end{picture}
\caption{ The $\sigma_j^x$--profile of the state $\mid \xi_{\alpha}\rangle$ near  the right
boundary for the choice  $\xi_{\alpha}=0.075, \xi_{\beta}=0.05$ in \eref{bx}.
The numerical data (+) have been obtained for $L=800$. The straight line
shows the behaviour according to \eref{bxr1}.
        }
\label{right}
\end{figure}

Our results  suggest that as long as $\xi_{\alpha}\neq \xi_{\beta}$
 the profile for the state $\mid\xi_{\alpha}\rangle$ 
is still given by \eref{bxl1} and \eref{bxr1}
(the deviation from the
expression in \eref{bxl1} for large $j$ in figure 1 is expected due to the presence of the right boundary).
 The profile for the state $\mid\xi_{\beta}\rangle$ 
(obtained by the excitation of the fermion with energy $2\Lambda_{\beta}=\cosh\xi_{\beta}$)
near the left boundary is just 
\begin{equation}
\langle \xi_{\beta} \mid \sigma_j^x \mid \xi_{\beta} \rangle \approx (-1)^j
Aj^{-\frac{1}{4}} 
\end{equation}
whereas at the right boundary we have
\begin{equation}
\label{bxr2}
\langle \xi_{\beta} \mid \sigma_{L+j-1}^x \mid \xi_{\beta} \rangle \approx (-1)^{j+1}
Aj^{-\frac{1}{4}}\left(1-2\rme^{(-1)^j\xi_{\beta}}\rme^{-2j\xi_{\beta}}\right).
\end{equation}
However, if $\xi_{\alpha}=\xi_{\beta}=\xi$ our computations suggest a different behaviour, i.e. 
\begin{equation}
\label{bxl3}
\langle \xi_{\alpha} \mid \sigma_j^x \mid \xi_{\alpha} \rangle=
\langle \xi_{\beta} \mid \sigma_j^x \mid \xi_{\beta} \rangle \approx
(-1)^{j}
Aj^{-\frac{1}{4}}\rme^{(-1)^j\xi}\rme^{-2j\xi}
\end{equation}
near the left boundary and 
\begin{equation}
\label{bxr3}
\langle \xi_{\alpha} \mid \sigma_{L+1-j}^x \mid \xi_{\alpha} \rangle=
\langle \xi_{\beta} \mid \sigma_{L+1-j}^x \mid \xi_{\beta} \rangle \approx
(-1)^{j}
Aj^{-\frac{1}{4}}\rme^{(-1)^j\xi}\rme^{-2j\xi}
\end{equation}
near the right boundary. In contrast to the profiles 
we have obtained so far, these profiles are left-right symmetric.
Furthermore the leading term proportional to $j^{-\frac{1}{4}}$
does not appear, unlike in equations \eref{bxl1} and \eref{bxr2}.

Note that the Hamiltonian with boundary conditions \eref{bx} conserves parity if
$\xi_{\alpha}=\xi_{\beta}$.
Our numerical computations yield that the energies $2\Lambda_{\alpha}$ and $2\Lambda_{\beta}$
are non-degenerate on the finite chain. Only in the limit $L\to \infty$ the degeneracy appears.
Hence the parity of the states $\mid \xi_{\alpha}\rangle$ and $\mid \xi_{\beta}\rangle$ 
is well-defined. This explains why the profiles reflect the symmetry of the Hamiltonian.

\section{Conclusions}
In this paper we have studied the one point functions of the $\sigma_j^z$ and $\sigma_j^x$ operator
for the XX-chain with hermitian boundaries, which in the continuum limit
 corresponds to a free boson field on a cylinder 
with Dirichlet-Dirichlet, Dirichlet-Neumann or Neumann-Neumann boundary conditions \cite{paper2}.

No results for the magnetization profiles in the ground-state  have
previously
 been obtained on the lattice.
The only known results  have been
found by means 
of the bosonization technique \cite{affleckx} and were restricted to
the limit $L\to \infty$ and $j\gg 1$ (see \eref{affx} and \eref{affz}).
The expression \eref{affz} is valid only for small values of the
longitudinal boundary field, i.e. for small values of $\alpha_z$ in our notation. 

We have obtained the exact ground-state profiles of the $\sigma_j^z$-operator  on the finite chain for the
boundary terms given in table 1.
The respective  profiles are given in \eref{exa}-\eref{exdeven}.
The continuum limit of the profiles is given in \eref{continuumz2+4}-\eref{sigmaz_case16_staggerd_Leven}.

The $\sigma_j^z$-profiles for the cases from \tref{paraz} in the limit $L\to \infty, j\gg 1$
are given in \eref{zt}.
Exact results for the $\sigma_j^x$-profiles in this limit  
have been found for the cases  in table 2 and are given in \eref{c2t}
 respectively \eref{cgt}. Our results obtained on the lattice agree to the expressions 
\eref{affx} and \eref{affz} which have been found using 
the bosonization technique \cite{affleckx}.
Note that in contrast to \eref{affz} 
our result \eref{zt} is not restricted to small values of the boundary field.

Supported by numerical computations, these exact results have led us to
conjectures for the profiles in the continuum limit 
 for more general values of the boundary parameters (see \eref{dd}-\eref{nn2}).
For the Dirichlet-Dirichlet case we have studied the most  general 
boundary terms \eref{dd}.
The $\sigma_j^z$-profile for these boundary terms is given in \eref{conconz}
(the $\sigma_j^x$ profile vanishes exactly, if the ground-state is non-degenerate).

We also considered a type of boundary terms, which correspond to the Dirichlet-Neumann boundary 
condition (see \eref{dn1}).
The $\sigma_j^x$ profile we conjectured is given in \eref{condnx}.
The profile of the $\sigma_j^z$ operator is given in \eref{condnz}.

The boundary terms given by \eref{nn1} and \eref{nn2} correspond to Neumann-Neumann 
boundary conditions on the boson field. 
The $\sigma_j^x$-profile for the first type of boundaries \eref{nn1} is given in \eref{connnx}.
Here the $\sigma_j^z$-profile vanishes exactly, if the ground-state is non-degenerate.
For the second type of boundary terms \eref{nn2} the profile of 
the $\sigma_j^z$ operator is given in \eref{connnz}. 
As in the case of the Dirichlet-Neumann boundary condition
 the $\sigma_j^x$ profile is not affected by the presence of the
diagonal boundary terms.

The expressions we have obtained for the profiles in the case of the Dirichlet-Dirichlet respectively
the Neumann-Neumann  boundary conditions do not coincide with the predictions 
\eref{scaling2} and \eref{ratioAB} conformal field theory makes for
 the profiles of  scaling operators near a boundary.
Here one has to consider certain linear combinations of spin-operators 
to obtain the correct behaviour. These combinations and
the respective behaviour near the boundary are given in \eref{combizprofil} and \eref{combixprofil}.

In this paper we have also discussed the appearance of massive excitations, i.e.
of fermions with a non-vanishing energy as $L$ goes to infinity.
Such a fermion appears if the reciprocal value of one of the expressions \eref{firstbound} or \eref{secondbound}
becomes larger than 1. These expressions reduce to \eref{bd} respectively \eref{bn}
in the case of purely non-diagonal respectively diagonal boundary terms.

We have computed the profiles near the boundaries
in the limit $L\to \infty, j\gg 1$
 for certain states with differ from the ground-state of $H_{\rm long}$ by 
 a massive excitation for
 boundary parameters satisfying \eref{bz} or \eref{bx}.
For the first type of boundary terms there exists one
massive excitation, whereas for the second type  there might appear
two of them.
The profiles near the boundaries are given by \eref{bzl},\eref{bzr}
respectively by \eref{bxl1}-\eref{bxr2}.
If the value of $\xi_{\alpha}$  equals $\xi_{\beta}$ (see \eref{bx})
the profiles are given by \eref{bxl3} and \eref{bxr3}.

\ack{The author is grateful to V.Rittenberg for many discussions
and useful suggestions. 
 }
\appendix
\section{Diagonalization of $H_{\rm long}$ and the projection mechanism}
\label{techsum}
In this appendix we are going to recapitulate the method we used in \cite{paper1} to
diagonalize $H_{\rm long}$ given in \eref{Hlong} and
the projection mechanism, which yields the eigenvalues and eigenvectors
of the Hamiltonian $H$ in \eref{HXX}. This is necessary in order to fix our notation  
we are going to use during our computations for the profiles of
$\sigma_j^z$ and $\sigma_j^x$  which are  presented in Appendix B respectively in Appendix C.

In its diagonal form $H_{\rm long}$ reads
\begin{equation}
H_{\rm long}=\sum_{k=0}^{L+1} 2\Lambda_k b_k a_k - \sum_{k=0}^{L+1}\Lambda_k
\label{H_long_diagonal}
\end{equation}
where the $b_k$ and $a_k$ satisfy the anti-commutation relations of
fermionic creation and annihilation operators.
The factor $2$ in equation \eref{H_long_diagonal} is just an remnant of
our notation introduced in \cite{paper1}.
The values of the $\Lambda_k$ are determined by the eigenvalues 
of a skew-symmetric $(2L+4)\times(2L+4)$ matrix $M$, i.e.
\begin{equation}
M(\phi_k^{\pm})=\pm\Lambda_k(\phi_k^{\pm}).
\label{M_eigenvalue_equation}
\end{equation}
The explicit form of $M$ 
 is  given in \cite{paper1}.
The eigenvalues $\Lambda_k$ can be obtained by the zeros $x_k$
of a polynomial $p(x^2)$, which has also been given in \cite{paper1}, i.e.
\begin{equation}
\Lambda_k=\frac{1}{4}(x_k+1/x_k) .
\end{equation}
We have also shown in \cite{paper1} how to compute
the eigenvectors $(\phi_k^{\pm})$ as a function of the $x_k$ 
and the boundary parameters $\alpha_{\pm}, \beta_{\pm}, \alpha_z$ and $\beta_z$.
For certain choices of the boundary terms 
the $x_k$ and the $(\phi_k^{\pm})$ can be computed exactly on the finite chain, since
the polynomial $p(x^2)$
factorizes into cyclotomic polynomials \cite{paper1}.
For the boundary parameters given in \tref{paraz} and \tref{parax} this is the case
\footnote{They correspond to the factorizable cases 2,4,9,11,14 and 16 in the notation of \cite{paper1}.}.
Otherwise the $x_k$ and the $(\phi_k^{\pm})$ may be computed 
solving \eref{M_eigenvalue_equation} numerically.

The eigenvectors $(\phi_k^{\pm})$ can be used to express the 
 Clifford operators $\tau_j^{\pm}$, which
are defined by 
\begin{equation} 
\tau_j^{+,-}=\left(\prod_{i=0}^{j-1}\sigma_i^z\right)\sigma_j^{x,y} 
\end{equation}
in the  terms 
 of the creation and annihilation operators $b_k$ and $a_k$, i.e.
\begin{equation}
\tau^{\mu}_j=\sum_{k=0}^{L+1} (\phi^-_k)^{\mu}_j a_k + (\phi^+_k)^{\mu}_j b_k 
\label{tauinfermion}
\end{equation}
where the $(\phi_k^{\pm})$ satisfy the orthogonality relation
 $\sum_{j,\mu} (\phi^+_l)^{\mu}_j (\phi^-_k)^{\mu}_j=\delta_{lk}$.
\Eref{tauinfermion} allows to write the spin operators in terms 
of the $b_k,a_k$ and the eigenvectors $(\phi_k^{\pm})$. 
Therefore the knowledge of these vectors is a crucial point 
in view to the computation of the magnetization profiles. 
 The vacuum state is defined by 
\begin{equation}
a_k|vac\rangle =0 \quad \forall k  .
\label{vacuum}
\end{equation}
Note, that the vacuum state itself is not element of one of the four sectors
$(+,+),(+,-),(-,-)$ or $(-,+)$. However, in view to applications to the Hamiltonian $H$ we are interested
into the eigenstates of $H_{\rm long}$ which lie in the $(+,+)$-sector.
Hence we have defined the states
\begin{equation}
\label{vpm}
|v^{\pm}\rangle = \frac{1}{\sqrt{2}}(1\pm b_0)|vac\rangle
\end{equation}
where $b_0$ denotes the creation operator of the spurious zero mode \cite{paper1} 
which is always present in the spectrum of
 $H_{\rm long}$.
The associated eigenvectors $(\phi_0^{\pm})$ are given by
\begin{equation}
\phi_0^+=(0,\frac{1}{2},0,\cdots,0,\frac{\rmi}{2},0) \qquad
\phi_0^-=(0,\frac{1}{2},0,\cdots,0,-\frac{\rmi}{2},0) 
\label{spm}
\end{equation}
where we used the notation
$(\phi_k^{\pm})=( (\phi_k^{\pm})_0^-,(\phi_k^{\pm})_0^+,(\phi_k^{\pm})_1^-,(\phi_k^{\pm})_1^+,\ldots,
(\phi_k^{\pm})_{L+1}^-, (\phi_k^{\pm})_{L+1}^+ )$.
The states $|v^{\pm} \rangle$ satisfy
\begin{equation}
\sigma_0^x |v^{\pm}\rangle = \pm |v^{\pm}\rangle  \qquad \sigma_{L+1}^x=\pm\eta |v^{\pm}\rangle
\end{equation}
where $\eta^2=1$.
If $\eta=1$ the state $|v^+\rangle$ is element of the $(+,+)$-sector and so are all
the states obtained by an even number of fermion excitations, where  we disregard 
the spurious zero mode.
 If $\eta=-1$ all the states obtained by the excitation of an odd number of fermions 
with respect to the state $|v^-\rangle$ are elements of the $(+,+)$-sector and hence
build up the spectrum of $H$, 
where we again have to disregard the spurious zero mode.
In the case of hermitian boundaries the value of $\eta$ may be computed via \cite{paper1}
\begin{equation}
\eta=(-1)^{L+1}\frac{\alpha_-\beta_++\alpha_+\beta_-}{4^{L+1}\prod_{k\neq 0} \Lambda_k}.
\label{eta}
\end{equation}
Note that \eref{eta} is of no use if there exist additional zero modes on top of the
spurious zero mode.

\section{Calculation of the $\sigma_j^z$ profiles}
\label{calproz}
In this appendix we are going to explain how to compute the $\sigma_j^z$-profiles
and how we have derived the ground-state profiles for the boundary terms in \tref{paraz}, i.e.
equations \eref{exa}-\eref{zt}.
We will also derive the profiles \eref{bzl} and \eref{bzr}
associated to the excitation of the massive fermion appearing 
for the boundaries \eref{bz}. Note that these boundaries coincide with the boundary terms 
of case e in \tref{paraz} if one identifies $\rme^{\xi}$ with $\tan(\frac{\pi}{4}+\frac{\delta}{2})$.

Using  \eref{tauinfermion}, \eref{vpm} and \eref{spm} one can show that 
 the $\sigma_j^z$ profile for $1\leq j\leq L$ and 
for any excited state $\mid \psi_r^{\pm} \rangle=\prod_{n=1}^{r}b_{k_n}\mid v^{\pm} \rangle$
with  $k_n\neq 0$
is given by 
\begin{equation}
\fl
\label{excz}
\langle \psi_r^{\pm}\mid \sigma_j^z \mid \psi_r^{\pm} \rangle=\langle vac\mid \sigma_j^z \mid vac \rangle
+\rmi\sum_{n=1}^r\left[ (\phi^-_{k_n})^+_j (\phi^+_{k_n})^-_j -(\phi_{k_n}^+)_j^+(\phi_{k_n}^-)_j^- \right] 
\end{equation}
where 
\begin{equation}
\label{szvac}
\langle vac \mid \sigma^z_j \mid vac \rangle  = - \rmi \langle vac \mid \tau^+_j \tau^-_j \mid vac \rangle
                                = - \rmi \sum_{k=0}^{L+1} (\phi^-_k)^+_j (\phi^+_k)^-_j.
\end{equation}
Before turning to the profiles for the boundaries in \tref{paraz},
consider  the
 case $\alpha_z=\beta_z=0$.
According to \cite{paper1} it is always possible to chose the eigenvectors $(\phi_k^{\pm})$ such
that the RHS of \eref{excz}
and \eref{szvac} vanishes for $0<j<L+1$ (see the equations (2.38) and (2.40) in \cite{paper1}).
In this case the profiles for all the states $\mid \psi_r^{\pm} \rangle$ vanish identically.
Thus the profile for the ground-state of $H$ vanishes also if it is non-degenerate. 
The partition functions obtained in \cite{paper2} show that for $L\gg 1$
this is the generic case. 
Therefore we considered only cases with at least one diagonal boundary term.
If the ground-state of $H$ is degenerate one has also to consider linear combinations 
of the states $\mid \psi_r^{\pm} \rangle$ which correspond to ground-states $H$.

For the cases given in \tref{paraz} 
 the polynomial $p(x^2)$ factorizes into
cyclotomic polynomials.
This enables us to compute the eigenvectors $(\phi_k^{\pm})$ exactly on the finite chain, 
as explained in \cite{paper1}.
The explicit computations  are cumbersome
but straight forward and will therefore not be given.

For all cases in \tref{paraz} there appears an additional zero mode on
top of the spurious mode which does not give a contribution to the RHS of 
\eref{excz}.
Hence the profiles for the two states $|v^+\rangle$ and $b_z|v^-\rangle$, where
 $b_z$  denotes the creation operator corresponding to the additional zero mode,
are identical. 
One of these states is element of the $(+,+)$ sector and corresponds to the
ground-state of $H$. However we do not have to know which one, since 
we are only interested into the profile (note that \eref{eta} is not applicable,
due to the zero mode).
Hence in the following we may assume 
 the state $|v^+\rangle$ to be in the $(+,+)$-sector.

For the cases a,b,c and d with $L$ even no complications arise.  
The ground-state of $H$ is non-degenerate and
using the expressions for the $(\phi_k^{\pm})$ in \eref{szvac} 
leads to a geometric series for each case.
Performing the computations yields \eref{exa}-\eref{exc} and \eref{exdeven}. 

For case d with $L$ odd the ground-state is two-fold degenerate,
since here two zero modes appear on top of the spurious zero mode.
The  non-vanishing components of the respective eigenvectors are given by
\begin{equation}
\label{z1}
(\phi_{z_1}^{\pm})_0^-=1 \qquad (\phi_{z_1}^{\pm})_{L+1}^+=\pm \rmi 
\end{equation}
respectively
\begin{equation}
(\phi_{z_2}^+)_j^-=-\rmi(\phi_{z_2}^+)_j^+ =
 \frac{(-\rmi)^j}{\sqrt{L}} \left\{ \begin{array}{ll} \rmi \quad & \mbox{for $j$ even} \\
                                        1 & \mbox{for $j$ odd}\end{array} \right.
\end{equation}
where $0<j\leq L$ and $(\phi_k^-)^{\pm}_j=\pm(\phi_k^+)_j^{\pm}$.
In this case we computed the profiles for the two states
 $| v^+ \rangle$ and $b_{z_2}b_{z_1}| v^+ \rangle$. This leads to \eref{exdodd} where
one has to choose the $-$ sign for the profile of $| v^+\rangle$ respectively 
the $+$ sign for the profile of $b_{z_2}b_{z_1}| v^+ \rangle$.

However, the computations for case e respectively the boundary terms \eref{bz}
are more involved. It is convenient to use the parameter $s=\rme^{-2\xi}$ 
where $\xi$ has been introduced in \eref{bz} instead of $\delta$ being used in \tref{paraz}.
For odd values of $L$  we have obtained
\begin{equation}
\langle \sigma_j^z \rangle = \frac{\varsigma_j}{\sqrt{s}L}-\frac{1-s}{1-s^L}s^{j-1}  .
\label{sigmaz_sum_case16_Lodd}
\end{equation}
where 
\begin{eqnarray}
\fl
\varsigma_j =\sum_{k=1}^{N}
\frac{x_k^{2j-3}+x_k^{-2j+3}-x_k^{2j-1}-x_k^{-2j+1} + s\left( x_k^{2j+1}+ x_k^{-2j-1}
- x_k^{2j-1}- x_k^{-2j+1}\right)}{1/s+s-x_k^2-1/x_k^2}
\label{varsigma}
\end{eqnarray}
with $x_k=\rme^{\rmi\frac{k\pi}{L}}$ and $N=(L-1)/2$.
This expression simplifies to a geometric series for $s=1$.  Straight forward
computation yields \eref{exdeven} as for case d with $L$ even. We have not been able to compute this expression
 for $s\neq 1$.

As for case d with $L$ odd, there appear two zero-modes on top 
of the spurious zero mode if we take $L$ even.
The non-vanishing components of the 
corresponding eigenvectors for the first zero mode are again given by \eref{z1} whereas for the 
second zero mode they read 
\begin{equation}
(\phi_{z_2}^+)_j^-=-\rmi(\phi_{z_2}^+)_j^+=\rmi^j \left( \frac{2}{L(1+s)}\right)^{1/2}
        \left\{ \begin{array}{rl} \rmi  \quad & \mbox{for $j$ even} \\
                                                -\sqrt{s} & \mbox{for $j$ odd} \end{array} \right.
\end{equation}
where $0<j\leq L$ and $(\phi_k^-)^{\pm}_j=\pm(\phi_k^+)_j^{\pm}$.
Again  we computed the profiles for the two states
 $| v^+ \rangle$ and $b_{z_2}b_{z_1}| v^+ \rangle$.
Labelling the profile for the state $| v^+ \rangle$ by a $-$ sign and
the profile for the state $b_{z_2}b_{z_1}| v^+ \rangle$ by a $+$ sign, we have obtained
\begin{equation}
\langle \sigma_j^z \rangle_{\pm} = \frac{\varsigma_j}{\sqrt{s}L} -\frac{1-s}{1-s^L}s^{j-1} \pm\frac{2}{L(1+s)}
\left\{ \begin{array}{ll}  1 \quad & \mbox{for $j$ even} \\
                           s & \mbox{for $j$ odd} \end{array} \right.
\label{sigmaz_sum_case16_Leven}
\end{equation}
where $\varsigma$ is again given by \eref{varsigma} as for $L$ odd, but with  $N=L/2-1$.
Choosing $s=1$ allows for the computation of the sum in \eref{varsigma} and yields \eref{exdodd}
as for case d with $L$ odd.

Although our results on the finite chain are restricted to $s=1$,
one can show that in the continuum limit ($z=j/L$ fixed, $L\to \infty$)  $\varsigma_j$  
is given by  
\begin{equation}
\varsigma_j= \frac{(-1)^j}{1/s+1}\frac{1}{\sin(\pi z)} 
\end{equation}
for $L$ odd respectively by
\begin{equation}
\varsigma_j= \frac{(-1)^j}{1/s+1}\cot(\pi z)
\end{equation}
for $L$ even.
Plugging this into \eref{sigmaz_sum_case16_Leven} and \eref{sigmaz_sum_case16_Lodd}
respectively results in the equations \eref{continuumz16odd} and
\eref{sigmaz_case16_staggerd_Leven}.
Note that the exponential contributions in  \eref{sigmaz_sum_case16_Lodd} and \eref{sigmaz_sum_case16_Leven}
vanish in the continuum limit.

The profile in the limit $L\to \infty,j$ fixed, leading to \eref{zt}, may also be computed.
In this limit
the quantity $\varsigma_j/L$ appearing in \eref{sigmaz_sum_case16_Lodd} and 
\eref{sigmaz_sum_case16_Leven}
 can be represented  
as  an  integral, i.e.
\begin{equation}
\fl
{\varsigma}_j^{\infty}=
\lim_{L\to\infty}
\frac{\varsigma_j}{L}=\frac{2}{\pi}\int\limits_{0}^{\frac{\pi}{2}}
\frac{\cos((2j-3)t)-(s+1)\cos((2j-1)t)+s\cos((2j+1)t)}{1/s+s-2\cos(2t)}\rmd t .
\end{equation}
This integral can be calculated exactly in terms of hypergeometric
functions. 
We obtained
\begin{eqnarray}
\fl
{\varsigma}_j^{\infty}=\frac{2}{\pi}\left[
\frac{(-1)^j}{2j-1}- (-1)^j(\frac{1}{s}-1)\frac{ F(1,1/2+j;3/2+j;-1/s)}{2j+1} \right]
\quad \mbox{for $s>1$} \\
\sim\frac{2}{\pi}\left[ \frac{(-1)^j}{j}\frac{s}{1+s}\right] + \Or(j^{-2})
\end{eqnarray}
and
\begin{eqnarray}
\fl
{\varsigma}_j^{\infty}=\frac{2}{\pi}\left[ \frac{(-1)^j}{2j-1}-(-1)^j(1-s)\frac{ F(1,1/2-j;3/2-j;-s)}{2j-1} \right]
\quad \mbox{for $0<s<1$} \\
\sim\frac{2}{\pi}\left[ \frac{(-1)^j}{j}\frac{s}{1+s}+\frac{\pi (1-s)}{2} s^{j-1/2}\right] + \Or(j^{-2}).
\label{uu}
\end{eqnarray}
For the asymptotic expansions of the hypergeometric functions see \cite{lukewatson}.
Using these results in \eref{sigmaz_sum_case16_Leven} and \eref{sigmaz_sum_case16_Lodd} 
gives for even and odd values of $L$
\begin{equation}
\label{gxi}
\langle \sigma_j^z \rangle = \frac{(-1)^j}{\pi \cosh \xi}j^{-1}+ \cdots
\end{equation}
Using $\rme^{\xi}=\sqrt{2}\alpha_z$ yields \eref{zt}.
Note that the exponential contributions in \eref{sigmaz_sum_case16_Leven} and \eref{sigmaz_sum_case16_Lodd}
(which, in the limit in question, are present only for $0<s<1$)
are  cancelled by the exponential part in \eref{uu}.

We have also considered the profile for a state 
where the massive fermion is excited, i.e.
 $\mid \xi \rangle= b_{\xi} b_z \mid v^+ \rangle$ for $L$ odd
respectively $\mid \xi \rangle= b_{\xi} b_{z_1} \mid v^+ \rangle$ for $L$ even 
(there are two zero modes for $L$ even, where the one labelled by $z_1$ gives
no contribution to the profile \eref{excz}).
Using \eref{excz},\eref{gxi} and the explicit expressions of the 
eigenvectors $(\phi_{\xi}^{\pm})$  corresponding to the massive fermion, i.e.
\begin{equation}
(\phi_{\xi}^+)_j^-=-\rmi (\phi_{\xi}^+)_j^+=
\rmi (-1)^j \left(\frac{1-\rme^{-2\xi}}{1-\rme^{-2\xi L}}\right)^{1/2}\rme^{-2\xi(j-1)}.
\label{bounds}
\end{equation}
where $0<j<L+1$ and $(\phi_{\xi}^-)^{\pm}_j=\pm(\phi_{\xi}^+)_j^{\pm}$, 
yields  \eref{bzl} for $\xi>0$ $(s<1)$.

Note that for $\xi<0$ $(s>1)$ and $j$ fixed the vector components \eref{bounds} vanish as $L$  
goes to infinity and the profile for the excited state $\mid \xi \rangle$ 
equals the ground-state profile given in \eref{gxi}.
By symmetry (see \eref{bz}) this yields the profile near the right boundary 
for the case $\xi>0$ $(s<1)$, i.e. equation \eref{bzr}.

The $U(1)$ charge, i.e. the projection of the total spin onto the z-axis,  
of the state $\mid \xi \rangle$ can be computed  using the 
explicit expressions of the $(\phi_k^{\pm})$ and
summing over $j$ in
\eref{excz} and \eref{szvac} before summing over $k$
(in this way the computation can be done exactly on the finite chain, independently of the
value of $s$).
This yields 0 for $L$ even respectively $\frac{1}{2}$ for $L$ odd as stated 
in section 5.2.

\section{Calculation of the $\sigma_j^x$-profiles}
\label{calprox}
In this appendix we are going to derive our exact results 
for the $\sigma_j^x$ profiles for the two cases 
from \tref{parax}, i.e. equations \eref{c2t},\eref{cgt} and \eref{sx9}.
Furthermore we are going to explain how we have obtained the  
profiles \eref{bxl1} and \eref{bxr1} 
corresponding to the boundary excitation
for the boundary terms in \eref{bx} with 
$\xi_{\beta}=-\xi_{\alpha}$.

We have shown in \cite{paper1} how to compute the expectation
values of the $\sigma_j^x$-operator in terms of pfaffians. 
Here we will only consider cases where diagonal boundary terms are absent
(see \tref{parax} and \eref{bx}). 
This enables us to represent $\langle \sigma_j^x\rangle$ in terms 
of the determinant of a $j\times j$ matrix \cite{paper1}, i.e.
\begin{equation}
\label{sigmajx}
\langle v^+ \mid \sigma_j^x \mid v^+\rangle = f_{0j} \left|
\begin{array}{cccc}
\langle \tau_0^- \tau_{1}^+ \rangle & \langle \tau_0^- \tau_1^- \rangle
 & \langle \tau_0^- \tau_3^+ \rangle & \cdots \\
\langle \tau_2^+ \tau_1^+ \rangle & \langle \tau_2^+ \tau_1^- \rangle
 & \langle \tau_2^+ \tau_3^+ \rangle & \cdots \\
\langle \tau_2^- \tau_1^+ \rangle & \langle \tau_2^- \tau_1^- \rangle
& \langle \tau_2^- \tau_3^+ \rangle & \cdots \\
\vdots & \vdots & \vdots & \ddots
\end{array}
\right|
\end{equation}
where $f_{0j}=-\rmi$ for odd values of $j$ and $f_{0j}=1$ for even values of $j$.
The expectation values of the basic contractions of pairs are given by 
\begin{equation}
\langle \tau_i^{\mu} \tau_j^{\nu} \rangle = \sum_{k=0}^{L+1} (\phi_k^-)_i^{\mu}(\phi_k^+)_j^{\nu}.
\label{basic_contractions}
\end{equation}
The profiles for the excited states  
\begin{equation}
\mid r \rangle =\prod_{j=1}^{r} b_{k_j} \mid v^{(-1)^r} \rangle
\end{equation}
are also given by the RHS of \eref{sigmajx} if one exchanges the $\langle \tau_i^{\mu} \tau_j^{\nu} \rangle$
in \eref{sigmajx} by
\begin{equation}
\label{bcex}
\langle \tau_i^{\mu} \tau_j^{\nu} \rangle_r =
\langle \tau_i^{\mu} \tau_j^{\nu} \rangle 
+\sum_{l=1}^r \left[(\phi_{k_l}^+)_i^{\mu}(\phi_{k_l}^-)_j^{\nu} - (\phi_{k_l}^-)_i^{\mu}(\phi_{k_l}^+)_j^{\nu}\right] .
\end{equation}
We will start with our computations for case g from \tref{parax} and will then turn to case f from \tref{parax}.
Afterwards we will consider the boundary terms \eref{bx}. 
Again it is possible to solve the eigenvalue equation \eref{M_eigenvalue_equation} exactly \cite{paper1}.
As for the case of the $\sigma_j^z$-profiles
we will not give the explicit expressions for the eigenvectors $(\phi_k^{\pm})$.
They have been obtained as explained in \cite{paper1}
\footnote{The boundary terms \eref{bx} correspond to case 10 in the notation of \cite{paper1}.}.

\subsection{Case g}
\label{caseg}
We will first show, how to derive equation \eref{sx9}.
Afterwards we will compute the function $f(j,L)$ in \eref{sx9} in the limit $L\to \infty, j\gg 1$.
Using \eref{sx9} this  yields
\eref{cgt}.

It turns out, that for our purpose it is convenient to consider the 
Hamiltonian $H_{\rm long}$ in a different basis, i.e.  we consider the
Hamiltonian
\begin{equation}
H'_{\rm long}=U_n H_{\rm long} U_n^{-1}
\end{equation}
where $U_n$ is given by 
\begin{equation}
U_n=
\left( \begin{array}{ll} 1 & 0 \\ 0 & 1 \end{array}\right)\otimes
\bigotimes_{j=1}^{L}\left(
\begin{array}{ll} 1 & 0 \\ 0 & \rme^{\rmi \varphi}\Gamma^j \end{array}\right)\otimes
\left( \begin{array}{cc} 1 & 0 \\ 0 & (-1)^n \end{array}\right)
\label{Trafo_spinbasis}
\end{equation}
and 
\begin{equation}
\Gamma = \exp\left(\rmi\frac{\chi+n\pi}{L+1}\right).
\end{equation}
The $\sigma_j^x$ profile for any state $\mid g\rangle$ in the original basis is related to the 
magnetization profiles
 for the state $\mid g'\rangle=U \mid g \rangle$ via
\begin{equation}
\label{sxg}
\fl
\langle g\mid \sigma_j^x \mid g \rangle=
 \cos\left(\varphi+ \frac{\chi+n\pi}{L+1}j \right) \langle g' \mid \sigma_j^x \mid g'\rangle 
+\sin\left(\varphi+ \frac{\chi+n\pi}{L+1}j \right) \langle g' \mid \sigma_j^y \mid g'\rangle .
\end{equation}
The Hamiltonian in the new basis can still be written in terms of free
fermions. 
Of course the fermion energies are not affected by the transformation.
The $L+1$ non-zero fermion energies are given by
\begin{equation}
\label{feg}
2\Lambda_k=\sin\left(\frac{2k+1}{L+1}\frac{\pi}{2} - \frac{\chi}{L+1} \right)
\end{equation}
where $1\leq k \leq L+1$. 
Computing the eigenvectors $(\phi_k^{\pm})'(\chi,\varphi)$ in the new basis yields that
they are independent from the values of $\chi$ and $\varphi$
and can be given in terms of the vectors $(\phi_k^{\pm})(0,0)$ in the original basis, i.e.
for $n\geq 0$ we have found
\begin{eqnarray}
\label{c11}
(\phi_k^{\pm})'(\chi,\varphi)=(\phi_{k+n}^{\pm})(0,0) \qquad &\mbox{for } 1\leq k \leq L+1-n \\
(\phi_k^{\pm})'(\chi,\varphi)=(\phi_{k+n-(L+1)}^{\mp})(0,0) \qquad & \mbox{for } L+1-n <  k \leq L+1
\end{eqnarray}
whereas for $n<0$ 
\begin{eqnarray}
(\phi_k^{\pm})'(\chi,\varphi)=(\phi_{k-|n|}^{\pm})(0,0) \qquad &\mbox{for } |n|+1\leq k \leq L+1 \\
(\phi_k^{\pm})'(\chi,\varphi)=(\phi_{L+1+j-|n|}^{\mp})(0,0) \qquad & \mbox{for } 1 \leq k < |n|+1 .
\label{c14}
\end{eqnarray}
Now define for $n$ fixed the state
\begin{equation}
\mid n \rangle = b_{L+2-n} \ldots b_{L+1} \mid v^{(-1)^n}\rangle \qquad \mbox{for } n>0
\end{equation}
respectively
\begin{equation}
\mid n \rangle = b_1 \ldots b_{|n|} \mid v^{(-1)^n}\rangle \qquad \mbox{for } n<0 .
\end{equation}
For $n=0$ define $\mid 0 \rangle=\mid v^+ \rangle$.
The expectation values in the new basis may be computed in the same way than in the original 
basis by exchanging the $(\phi_k^{\pm})$ in \eref{bcex} and \eref{basic_contractions}
by the $(\phi_k^{\pm})'$.
Using  \eref{c11}-\eref{c14} yields that the 
$\sigma_j^x$-profiles for the states $\mid n \rangle$ in the new basis are given by 
\begin{equation}
\label{sx}
\langle n '\mid \sigma_j^x \mid n' \rangle=\langle v^+ \mid \sigma_j^x \mid v^+ \rangle .
\end{equation}
where the RHS has to be evaluated in the old basis and for the choice $\chi=\varphi=0$.
In the same way one can show that 
\begin{equation}
\label{sy}
\langle n '\mid \sigma_j^y \mid n' \rangle=\langle v^+ \mid \sigma_j^y \mid v^+ \rangle
\end{equation}
where again the RHS has to be computed in the original basis for the choice $\chi=\varphi=0$.
Since for $\chi=\varphi=0$ the Hamiltonian $H_{\rm long}$ commutes not only 
with $\sigma_0^x$ and $\sigma_{L+1}^x$, but also with $\sigma_0^x\sigma_1^x\cdots \sigma_{L+1}^x$,
the $\sigma_j^y$ profile in \eref{sy} vanishes. 

However, we have been interested in the ground-state profiles.
Computing the value of $\eta$ using \eref{eta} with \eref{feg} yields $\eta=(-1)^{L+1}$
and hence the state $\mid n \rangle$ is element of the $(+,(-1)^{L+1+n})$-sector
(the excitation of a fermion changes the sector from $(\pm,\epsilon)$ to
$(\mp,\epsilon)$ \cite{paper1}).
Which of states $\mid n\rangle$ corresponds to the ground-state of $H$ may be
determined by computing the associated energies  using \eref{feg}.

For $L$ odd  the ground-state of $H$ corresponds 
to the state $\mid 0 \rangle= \mid v^+ \rangle$ if $\chi\neq \pi$.
For $\chi=\pi$ the ground-state of $H$ is twofold degenerate.
We obtain two linear independent ground-states by choosing the states $\mid 0\rangle$ and
$\mid -2\rangle$.

For $L$ even and $\chi >0$ the ground-state of $H$ corresponds to $\mid -1\rangle$ 
whereas for $\chi<0$ the ground-state corresponds to $\mid 1 \rangle$.
For $\chi=0$ the ground-state is twofold degenerate and we obtain
two linear independent ground-states choosing the states $\mid \pm 1\rangle$.
Finally, using \eref{sxg} and \eref{sx}
 with $f(j,L)=\langle v^+ \mid \sigma_j^x \mid v^+ \rangle$ (for the choice $\chi=\varphi=0$)
  and  $m=n/2$  
yields \eref{sx9} and the values of $m$ as explained in the text.

 In order to compute $f(j,L)$ we have calculated the expectation values of the
basic contractions of pairs \eref{basic_contractions}
using the vectors $(\phi_k^{\pm})(0,0)$.
For $i,j<L+1$ this results in
\begin{equation}
\langle\tau_i^{\pm}\tau_j^{\pm}\rangle = 0 \qquad \langle\tau_0^-\tau_j^+\rangle =
\sqrt{2} g(j)
\label{basiccontractions_case9_1}
\end{equation}
\begin{equation}
\langle\tau_i^+\tau_j^-\rangle =
g(i+j)-g(i-j)\qquad
\langle\tau_i^-\tau_j^+\rangle =
g(i+j)+g(i-j)
\label{basiccontractions_case9_2}
\end{equation}
where the values of $i$ are always even and the values of $j$ are always odd.
Furthermore we have defined
\begin{equation}
g(r)=\frac{-\rmi}{L+1}\frac{\sin(r\pi/2)}{\sin(r\pi/(2L+2))}.
\label{g}
\end{equation}
Using elementary operations on determinants in \eref{sigmajx} one can show that
\begin{equation}
\label{fjL}
f(j,L) = (-1)^j \left(\frac{2}{\pi}\right)^j\sqrt{2} \det{\mbox{\sf Q}}_j  ,
\end{equation}
where the matrix elements of the $j\times j$-matrix $\mbox{\sf Q}_j$ are given by
\begin{equation}
\label{qjkl}
({\mbox{\sf Q}}_j)_{kl}=\left(\frac{\pi}{2L+2}\right)^j  \left(\sin\frac{(2k-2l+1)\pi}{2L+2}\right)^{-1} .
\end{equation}
The boundary behaviour of the profile (i.e. \eref{cgt}) is obtained taking 
the limit $L\to \infty$ of ${\mbox{\sf Q}}_j$, which yields
\begin{equation}
\lim_{L\to \infty}
{\mbox{\sf Q}}_j=\left(
\begin{array}{ccccc}
1 & -1 & -\frac{1}{3} & \ldots & \frac{1}{3-2j} \\
\frac{1}{3} & 1 & -1 & \ldots & \frac{1}{5-2j} \\
\frac{1}{5} & \frac{1}{3} & 1 & \ldots & \frac{1}{7-2j} \\
\vdots & \vdots & \vdots & \ddots & \vdots \\
\frac{1}{2j-1} & \frac{1}{2j-3} & \frac{1}{2j-5} & \ldots & 1
\end{array}
\right) \quad.
\end{equation}
The large $j$  behaviour of the determinant of this matrix is already known \cite{mccoywu}.
With \eref{fjL}
we obtain for $j\gg 1$
\begin{equation}
f(j,\infty) = (-1)^j  A j^{-\frac{1}{4}} +\cdots
\label{leftboundary_case9}
\end{equation}
where $A$  is defined by \eref{lnA}.
Using \eref{sx9} this  yields
\eref{cgt}.
The function $f(j,L)$ may also be calculated on the finite chain using 
 \eref{qjkl} in \eref{fjL} and computing the determinant numerically.

\subsection{Case f} 
\label{casef}
For this case there exists an additional zero mode in the spectrum.
Hence the ground-state of $H$ corresponds to $\mid v^+ \rangle$ or to
$b_z\mid v^- \rangle$, but we can not decide which one, since \eref{eta}
can not be applied.
However,
the eigenvectors $(\phi_z^{\pm})$ corresponding to the 
additional zero mode
give no contribution to \eref{bcex} as long as $i,j< L+1$ .
Hence the profiles for states $\mid v^+ \rangle$ and $b_z\mid v^- \rangle$
 are identical for $j<L+1$.  
Defining 
\begin{equation}
f(r)=\cos\left(\frac{r}{2L+2}\pi\right)g(r)
\end{equation}
where $g(r)$ is already given in \eref{g},
we have obtained for $L$ odd:
\begin{equation}
\langle\tau_i^{\pm}\tau_j^{\pm}\rangle = 0 \qquad 
\langle\tau_0^-\tau_j^+\rangle = \sqrt{2} g(j)
\label{bc1case2Lodd}
\end{equation}
\begin{equation}
\langle\tau_i^+\tau_j^-\rangle = 
f(i+j)-f(i-j) \qquad
\langle\tau_i^-\tau_j^+\rangle =
g(i+j)+g(i-j)
\end{equation}
where $i$ is always even and $j$ is always odd.
For $L$ even we obtain 
\begin{equation}
\langle\tau_i^{\pm}\tau_j^{\pm}\rangle = 0 \qquad
\langle\tau_0^-\tau_j^+\rangle =\sqrt{2} f(j)
\end{equation}
\begin{equation}
\langle\tau_i^+\tau_j^-\rangle =
g(i+j)-g(i-j) \qquad
\langle\tau_i^-\tau_j^+\rangle =f(i+j)+f(i-j) .
\label{bc2case2Leven}
\end{equation}
The expressions \eref{bc1case2Lodd}-\eref{bc2case2Leven}
have the same large $L$ limit as
 the respective expressions for case g in \eref{basiccontractions_case9_1} and
\eref{basiccontractions_case9_2}.
Thus the behaviour of the ground-state profile near the left
boundary is also given by the RHS of equation \eref{leftboundary_case9} which 
yields \eref{c2t}.

\subsection{ }
\label{fuck}
Here we consider the $\sigma_j^x$-profile for the boundaries 
in \eref{bx} with $\xi_{\beta}=-\xi_{\alpha}$. 
In this case there appears a massive fermion with energy $2\Lambda=2\cosh\xi_{\alpha}$ (see section 5).
Here we are  interested into the effect 
of the excitation of this fermion onto the magnetization profile.
For simplicity we use $\xi\equiv \xi_{\alpha}$ in the following.

We have restricted ourselves to even values of
$L$. 
Since $\eta=(-1)^{L+1}$ the state $|\xi \rangle=b_{\xi} \mid v^-\rangle$
(where by $b_{\xi}$ we denote the creation operator associated to the massive fermion)
lies in the $(+,+)$-sector of the Hilbert-space of $H_{\rm long}$ and hence corresponds to an eigenstate of $H$. 
For odd values of $L$ the $(+,+)$-sector is given by the 
excitations of an even number of fermions with respect to the ground-state $| v^+ \rangle$
and hence the state $|\xi \rangle$ does not correspond to an eigenstate
of $H$ for $L$ odd.

A lengthy computation yields the 
 expectation values of the 
basic contractions of pairs \eref{basic_contractions}, i.e.
\begin{equation}
\langle \tau_i^+\tau_j^- \rangle= 
g(i+j)-g(i-j)
\end{equation}
\begin{equation}
\langle \tau_i^-\tau_j^+ \rangle =\rmi\frac{\Xi(i+j)+1}{L+1}
+g(i-j)
+\frac{2\rmi\sinh(2\xi)\rme^{-(i+j)\xi}}{\rme^{-2(L+1)\xi}-1}
\end{equation}
\begin{equation}
\langle \tau_0^-\tau_j^+ \rangle=
\frac{\sqrt{2}\alpha_+}{\rme^{2\xi}+1}\left( \rmi\frac{\Xi(j)+1}{L+1} +g(j)
+\frac{2\rmi\sinh(2\xi)\rme^{-j\xi}}{\rme^{-2(L+1)\xi}-1} \right)
\end{equation}
where $g(r)$ has been defined in \eref{g} and
\begin{eqnarray}
\fl
\Xi(q)=
\sum_{n=1}^{\frac{L}{2}}
\left[\rme^{2\xi}\cos\left(\frac{q-2}{L+1}n\pi\right)+
      \rme^{-2\xi}\cos\left(\frac{q+2}{L+1}n\pi\right)-
      2\cos\left(\frac{q}{L+1}n\pi\right)\right] \\
\times
\left[ \cosh(2\xi)-\cos\left(\frac{2n}{L+1}\pi\right)\right]^{-1} .
\label{Xi}
\end{eqnarray}
For $L \to \infty$ we can rewrite $\frac{\Xi(q)}{L+1}$ in terms of an integral
which can be calculated using standard methods.
In this limit we may write
\begin{equation}
\label{c1}
\langle \tau_i^+\tau_j^- \rangle = -\frac{2\rmi}{\pi}\left( \frac{\sin((i+j)\pi/2)}{i+j}
                                                         -  \frac{\sin((i-j)\pi/2)}{i-j}\right)
\end{equation}
\begin{eqnarray}
\langle \tau_i^-\tau_j^+ \rangle =& -\frac{2\rmi}{\pi} \Bigg[
\frac{\sin((i-j)\pi/2)}{i-j}+\frac{\sin((i+j)\pi/2)}{i+j}\rme^{2\xi}\nonumber \\
&-2\sinh(2\xi)\rme^{-(i+j)\xi} \Bigg( \sum_{m=0}^{\frac{i+j-1}{2}}\frac{(-1)^m \rme^{(2m+1)\xi}}{2m+1}
- \arctan(\rme^{\xi}) \Bigg) \Bigg]
\end{eqnarray}
\begin{eqnarray}
\label{c3}
\langle \tau_0^-\tau_j^+ \rangle =& -\frac{2\sqrt{2}\alpha_+\rmi}{\rme^{2\xi}+1} \Bigg[
\frac{\sin(j\pi/2)}{j}(1+\rme^{2\xi}) \nonumber \\
& -2\sinh(2\xi)\rme^{-j\xi}  \Bigg(  \sum_{m=0}^{\frac{j-1}{2}}\frac{(-1)^m \rme^{(2m+1)\xi}}{2m+1}
- \arctan(\rme^{\xi}) \Bigg) \Bigg].
\end{eqnarray}
We used these expressions for the numerical calculation of 
the determinant in \eref{sigmajx} for $j=1,2\ldots 100$.
We have chosen 48 different values of $\xi$ in the range between $-4.0$ and $4.0$.
Numerical extrapolation yields 
\begin{equation}
\langle v^+\mid \sigma_j^x \mid v^+ \rangle \sim  (-1)^j A j^{-1/4}
\end{equation}
where the numerical value of $A$ is given by the RHS of \eref{C}
with an accuracy of the order of $10^{-13}$.

As already mentioned above here we are interested into  the profile of the state where the massive fermion
 is excited with respect to $|v^-\rangle$.
The eigenvectors $(\phi_{\xi}^{\pm})$ associated to the massive fermion are given by
\begin{equation}
\label{e1}
(\phi_{\xi}^+)_0^-=\frac{-\alpha_+}{\cosh\xi}/\rho
 \quad (\phi_{\xi}^+)_0^+=0
\label{boundxi1}
\end{equation}
\begin{equation}
(\phi_{\xi}^+)_j^-=\left\{
\begin{array}{ll}
-\sqrt{2}\exp((1-j)\xi)/\rho \quad & \mbox{for $j$ even} \\
0 & \mbox{for $j$ odd}
\end{array}\right.
\end{equation}
\begin{equation}
(\phi_{\xi}^+)_j^+=\left\{
\begin{array}{ll}
0 & \mbox{for $j$ even} \\
\rmi \sqrt{2} \exp((1-j)\xi)/\rho \quad & \mbox{for $j$ odd}
\end{array}\right.
\end{equation}
\begin{equation}
(\phi_{\xi}^+)_{L+1}^-=0 ,\quad
(\phi_{\xi}^+)_{L+1}^+=\frac{\rmi \beta_+}{\cosh\xi}\exp((1-L)\xi)/\rho
\label{boundxi2}
\end{equation}
where
\begin{equation}
\label{normxi}
\rho=\sqrt{\frac{\rme^{-2L\xi}-\rme^{2\xi}}{\sinh(2\xi)}} 
\end{equation}
and $(\phi_k^-)_j^{\mu}=(-1)^j(\phi_k^+)_j^{\mu}$.
In the case $\xi <0$ this mode does not contribute to 
\eref{bcex} in the limit $L\to \infty$, since
 \eref{normxi} diverges, i.e.
\begin{equation}
\label{xik0}
\langle \xi \mid \sigma_j^x \mid \xi \rangle \sim  (-1)^j A j^{-1/4} \quad \mbox{for } \xi<0 .
\end{equation}
This changes for $\xi >0$.
We assumed the following behaviour of the profile:
\begin{equation}
\langle \xi \mid \sigma_j^x \mid \xi \rangle \sim 
(-1)^{j+1} A(1-c\rme^{-j\eta})j^{-1/4} .
\label{boundprof}
\end{equation}
Using \eref{c1}-\eref{c3} and \eref{e1}-\eref{normxi} in  \eref{bcex} 
we computed the profile for $j=1,2...100$ for $24$ different values of $\xi$ 
in the range between $0$ and $4$.
The values of $c$ and $\eta$ have been computed using numerical extrapolation by
considering the quantity 
\begin{equation}
\ln\left(1+\frac{\langle \xi \mid \sigma_j^x \mid \xi \rangle}
 	{\langle v^+ \mid \sigma_j^x\mid v^+ \rangle} \right) \sim  \ln c -j\eta 
\end{equation}
Our results have confirmed our assumption \eref{boundprof}  and the numerical estimates
from our extrapolations suggest
\begin{equation}
c=2\rme^{(-1)^j \eta} \quad  \eta=2\xi 
\label{boundconst}
\end{equation}
which yields \eref{bxl1}.
The largest deviation of the extrapolated values for $c$ and $\eta$ from \eref{boundconst}
has been of the order of $10^{-11}$. We omit any tables on this.
Computing the profile for $\xi>0$ near the right boundary is equivalent to
the computation of the profile near the left boundary for $\xi<0$ (see \eref{bx})
which is given by \eref{xik0}. This yields \eref{bxr1}.

\section{Numerical verification of the conjectures in section 3}
\label{cnum}
We have verified our conjectures \eref{conconz}-\eref{connnz} for the continuum limit
of the magnetization profiles for the boundary terms \eref{dd}-\eref{nn2}.
In order to do so, we have used numerical diagonalization 
to obtain the  vectors $(\phi_k^{\pm})$
  (see \eref{M_eigenvalue_equation}) numerically
and used  \eref{excz}  and \eref{sigmajx} to obtain the 
profiles of $\sigma_j^z$ respectively $\sigma_j^x$ on the finite chain. 
Notice that \eref{sigmajx} is only valid if $\alpha_z=\beta_z=0$ and 
can not be applied to compute the $\sigma_j^x$ profile
for the boundary terms \eref{nn2}.
In this case we made use of the determinant representation (10.14) in \cite{paper1} of
the pfaffian which determines the profiles.

We computed the profiles on finite lattices with $L=20,60,100,\ldots$
 respectively $L=21,63,105,\ldots$, where the maximum value of $L$
has been chosen in the range of $500-700$.
For each type of boundaries in \eref{dd}-\eref{nn2}
 we have used $30-50$ different values of the respective free
parameters.

The values obtained on the  finite lattice have been extrapolated 
to $L=\infty$ for $z=j/L$ fixed using the BST algorithm, where
 $z=\frac{n}{20}$ for $L$ even respectively $z=\frac{n}{21}$ for $L$ odd  with $n\in \mathbb{N}$.

\begin{figure}[tbp]
\setlength{\unitlength}{1mm}
\def\setl{ \setlength\epsfxsize{8.0cm}}
\begin{picture}(155,90)
\put(28,80){
        \makebox{
                \setl
                \epsfbox{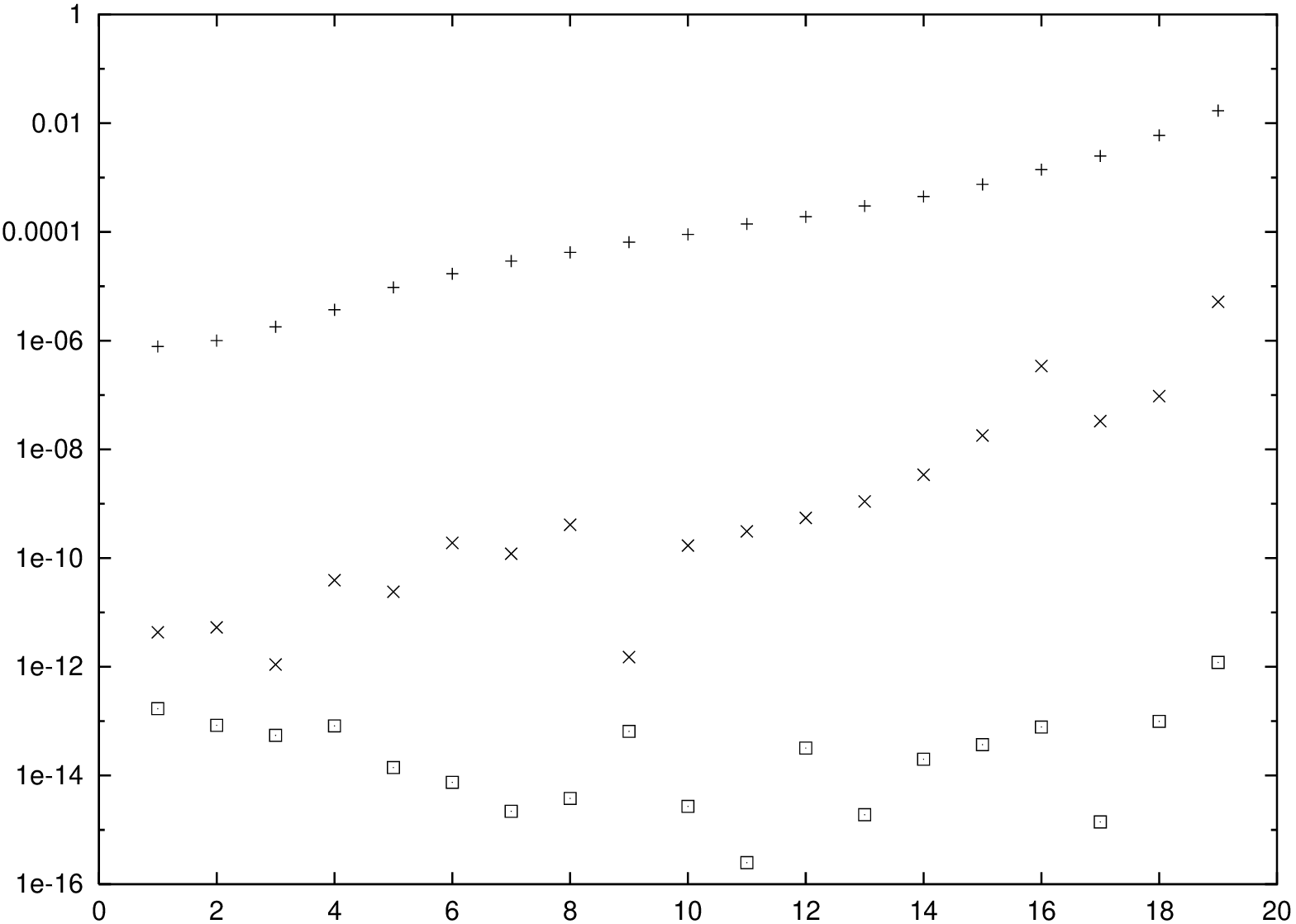}}
        }
\put(10,81){\footnotesize $\left| 1-\langle \sigma_j^x \rangle_{\rm extrapol}\big/\langle \sigma_j^x \rangle_{\rm conjecture} \right|$ }
\put(48,70){\footnotesize $\beta_{\pm}=\sqrt{8}/100$ }      
\put(72,70){\footnotesize $+$} 
\put(48,64){\footnotesize $\beta_{\pm}=\sqrt{8}/20$}
\put(70,64){\footnotesize $\times$}
\put(48,58){\footnotesize $\beta_{\pm}=\sqrt{8}$}
\put(66,58){\footnotesize $\Box$}
\put(92,-2){$z$}
\end{picture}
\caption{The absolute value of the relative deviation of the extrapolated profile $\langle \sigma_j^x \rangle_{\rm extrapol}$
from our conjecture \eref{condnx} for the boundaries \eref{dn1}.
The extrapolated profiles have been obtained from finite size data
 for lattice lengths $L=20,60,\ldots,700$.
        }
\label{reldev}
\end{figure}

As already mentioned in the text, the relative deviation of our 
numerical results from the conjectures is typically of the order of $10^{-12}-10^{-7}$.
In general the accuracy of our computations depends
on the choice of the parameters in  \eref{dd}-\eref{nn2}
and we have searched for regions in the parameter space where the accuracy 
of our computations breaks down.
We found that this is the case for small values of $\alpha_{\pm}$ and $\beta_{\pm}$ in \eref{dn1} 
respectively in \eref{nn1} and \eref{nn2}.
As an example \fref{reldev} shows the relative deviation
of the extrapolated $\sigma_j^x$-profile for the Dirichlet-Neumann boundary 
condition \eref{dn1} with $\alpha_z=0$  from the 
exact expression \eref{condnx} for $\beta_{\pm}=\sqrt{8}, \beta_{\pm}=\sqrt{8}/20$, $\beta_{\pm}=\sqrt{8}/100$
and for even values of $L$, where we have used finite size data from $L=20$ up to $L=700$.
The deviations are always in the range of the accuracy of the extrapolations. Hence we avoid
any error bars.

The behaviour illustrated in  \fref{reldev} is not unexpected, since the vanishing 
of a non-diagonal boundary term implies a sudden change of the respective boundary condition 
on the free boson from Neumann to Dirichlet. 
For the boundaries in question this change implies that the $\sigma_j^x$ profile vanishes exactly
even on the finite chain
(the Hamiltonian $H$ commutes with $\sigma_1^z\sigma_2^z\cdots\sigma_L^z$ for $\beta_{\pm}=0$
 and the ground-state of $H$ is non-degenerate).
 Since the profiles on the finite chain 
are smooth functions of all parameters appearing in the Hamiltonian, the deviation 
of the finite size data from the profile \eref{condnx} we conjectured for the case $\beta_{\pm}\neq 0$ 
in the continuum limit  (which
does not depend on the value of $\beta_{\pm}$)
increases as $\beta_{\pm}$ goes to zero. Assuming that our conjecture is correct,
 it is not surprising that the accuracy of our extrapolations decreases 
simultaneously.

\end{document}